\documentstyle[onecolumn,psfig]{mn}

\title[Universal Galaxy Rotation and Dark Matter]
{\bf THE UNIVERSAL ROTATION CURVE OF SPIRAL GALAXIES:
I. THE DARK MATTER CONNECTION} 
\author[M.Persic, P.Salucci and F.Stel ]
{Massimo Persic$^{1,2}$, Paolo Salucci$^{2}$, and Fulvio Stel$^{3}$ \\
$^1$ Osservatorio Astronomico, via G.B. Tiepolo 11, I-34131 Trieste, Italy   \\
$^2$SISSA -- International School for Advanced Studies,
via Beirut 4, I--34013 Trieste, Italy \\
$^{3}$Dipartimento di Astronomia, Universit\`a di Trieste,
via Beirut 4, I--34013 Trieste, Italy \\\
persic@tsmi19.sissa.it, salucci@tsmi19.sissa.it, fulvio@tsmi19.sissa.it}

\begin{document}

\maketitle

\def\mincir{\raise -2.truept\hbox{\rlap{\hbox{$\sim$}}\raise5.truept
\hbox{$<$}\ }}
\def  \magcir{\raise -2.truept\hbox{\rlap{\hbox{$\sim$}}\raise5.truept
\hbox{$>$}\ }}

\def\ref{\par\noindent\hangindent 20pt}
\def \dvd{\nabla_d}
\def \dvh{\nabla_h}
\def \dv{\delta V}
\def \dvlum{\delta V_{lum}}

\def \Ro{R_{opt}}

\begin{abstract}

We use a homogeneous sample of about 1100 optical and radio rotation curves (RCs) 
and relative surface photometry to investigate the main mass structure properties 
of spirals, over a range of 6 magnitudes and out to $\mincir 1.5$ and 2 optical 
radii (for the optical and radio data, respectively). We definitely confirm the 
strong dependence on luminosity for both the {\it profile} and the {\it amplitude} 
of RCs claimed by Persic \& Salucci (1991). Spiral RCs show the striking feature 
that a single global parameter, e.g. luminosity, dictates the rotation velocity at 
any radius for any object, so unveiling the existence of a Universal RC. At high 
luminosities, there is a slight discrepancy between the profiles of RCs and those 
predicted from the luminous matter (LM) distributions: this implies a small, yet 
detectable, amount of dark matter (DM). At low luminosities, the failure of the LM 
prediction is much more severe, and the DM is the only relevant mass component. We 
show that the Universal RC implies a number of scaling properties between dark and 
luminous galactic structure parameters:
{\it (a)} the DM/LM mass ratio scales inversely with luminosity; 
{\it (b)} the central halo density scales as $L^{-0.7}$;
{\it (c)} the halo core radius is comparable to the optical radius, but shrinks 
for low luminosities; 
{\it (d)} the total halo mass scales as $L^{0.5}$.
Such scaling properties can be represented as a curve in the (luminosity)-(DM/LM 
mass ratio)-(DM core radius)-(DM central density) space, which provides a 
geometrical description of the tight coupling between the dark and the luminous 
matter in spiral galaxies. 

\end{abstract}

\section{Introduction}

It is well known that rotation curves (hereafter RCs) of spiral galaxies do not
show any Keplerian fall-off which implies, as a most natural explanation, the
presence of an additional and invisible mass component (Rubin et al. 1980; Bosma 
1981b; Faber \& Gallagher 1979; see also Ashman 1992). In more detail, the profiles 
of the RCs imply that the distribution of light does not match the distribution of 
mass: in fact, in each galaxy the {\it local} $M/L$ ratio increases with radius by 
up to a factor $\sim 10^3$. Moreover, decreasing the galaxy luminosity, the light
is progressively more unable to trace the radial distribution of the dynamical mass
(e.g.: Persic \& Salucci 1988, 1990a,b, 1991; Broeils 1992b). This is evident in
the {\it Universal Rotation Curve} of spirals, pioneered by Rubin and
collaborators (e.g., Rubin et al. 1985; see also Burstein \& Rubin 1985 for the
related concept of 'mass types') and derived by Persic \& Salucci (1991;
hereafter PS91) from a sample of 58 late-type spiral RCs. PS91 found that
$$
V(R) ~ \simeq ~ 200 ~ \biggl( {L \over L_*} \biggr)^{\alpha}~
           [1 + f(L_B/L_B^*)~ ({R \over R_M} -1)] ~~{\rm km ~s}^{-1}
	~~~~~~~~~ 0.5 \mincir R/R_{opt} \mincir 1
\eqno(1)
$$
with $\alpha \simeq 0.25$, $R_M = 2.2\,R_D$ \footnote{$R_D$ is the disc exponential 
length-scale; $R_{opt} \equiv 3.2\,R_D$ is the optical radius (see below).}, 
log$L_B^*=10.4$ ($L_B$ denotes $B$ luminosities)\footnote{A value of the Hubble 
costant of H$_0=75$ km s$^{-1}$ Mpc$^{-1}$ is used throughout this paper.}, and
$f(L_B/L_B^*)$ a function linear in log$(L_B/L_B^*)$ which takes the values of 
0.5 and 0 at log$(L_B/L_B^*) = -1.2$ and 0.4, respectively. The main features of
eq.(1) are: {\it (a)} the linear dependence on radius, and {\it (b)} the luminosity 
dependence of the slope 
$$
\nabla ~\equiv~ {dV \over dR} {R \over V}\bigr|_{R_{opt}} ~=~ 
g(L_B/L_B^*)
$$
where $g(L)$ is a decreasing function of luminosity ranging between 
$-0.1 $ and $ 0.6$ (see PS91). A similar luminosity dependence of the RC shapes is 
seen also at outer radii (i.e., for $R_{opt} \mincir R \mincir 2\,R_{opt}$): the 
RCs of low-luminosity galaxies continue to rise and then flatten out 
asymptotically, while the RCs of high-luminosity objects show a distinctive drop 
just outside $R_{opt}$, followed by an asymptotically flat regime with $V_{\infty} 
\leq V(R_{opt}$) (Salucci \& Frenk 1989; see also Casertano \& van Gorkom 1991). 

This dependence of the RC shapes has a pivotal importance in the dark matter (DM) 
issue because it implies that the structural properties of dark and luminous matter 
are connected (see Appendix A), probably as a consequence of the process of galaxy 
formation itself. In addition, the study of RC profiles can provide a 
straightforward test for power-law perturbation spectra (Moore 1994), and so it is 
a crucial aspect in formulating theories of galaxy formation (e.g., Evrard et al. 
1994; Navarro \& White 1994; Flores et al. 1993; Navarro, Frenk \& White 1996). 

On the other hand, over the last few years the amount of available data has 
increased by more than one order of magnitude (e.g., Broeils 1992b; Amram et al.
1992, 1994; Schommer et al. 1993; Persic \& Salucci 1995, hereafter PS95). For
instance, the PS95 sample (967 curves) is a factor $\sim$15 bigger than that used 
in PS91. It is worthwhile, then, to make a definitive assessment of our previous
results (e.g., Persic \& Salucci 1988, 1990b, 1991) and of their interpretation in
the light of the recent theoretical work. In detail, the aim of this paper is to 
derive the {\it Universal Rotation Curve} of spiral galaxies and to investigate 
the main properties of the dark matter distribution. 

The plan of this paper is the following: in Section 2 we describe the selection
criteria and the procedure used to build the RC samples; in Section 3 we introduce 
two model-independent DM indicators and we study the DM systematics; in Section 4 
we derive the Universal Rotation Curve; in Section 5 we discuss some implications 
of our results for current scenarios of galaxy formation. Finally, in Section 6 we 
comment on our main results and draw conclusions.

\section{Selection Criteria and RC Samples}

In order to evaluate the ability of a RC to probe the distribution of matter out to 
the outskirts of the distribution of the luminous matter (LM), we set a suitable 
reference scale: $R_{opt}$, the radius encompassing 83\% of the total integrated 
light. For an exponential surface brightness distribution, $I(r) \propto 
e^{-r/R_D}$, we have $R_{opt}=3.2\,R_D$\footnote{ For a Freeman disc this 
corresponds to the de Vaucouleurs 25 $B$-mag/arcsec$^2$ photometric radius.}. 
Therefore, we can consider that the stellar disc has a size $\simeq R_{opt}$ and 
that no appreciable amount of LM exists farther out. Obviously a RC not extended 
out to $R_{opt}$ gives little information on the underlying gravitational potential 
(see Lake \& Feinswog 1989). Let us now introduce the samples of RCs: 
\bigskip

\noindent
$\bullet$ {\it Sample A.} It includes 131 individual RCs with a reliable profile
out to $R_{opt}$. To ensure this, we have set the following selection criteria.
In order to be included, an RC must: {\it (a)} extend out to $\geq R_{opt}$;
{\it (b)} have at least 30 velocity measurements distributed homogeneously with
radius and between the two arms; and {\it (c)} show no global asymmetries or
significant non-circular motions: the profiles of the approaching and receding
arms must be compatible. For 21-cm RCs we additionally require that {\it (d)}
the beam-size be $\leq 1/3\,R_{opt}$. Let us notice that criterion {\it (b)} is
also required to apply criterion {\it (c)}. 
An RC fulfilling these requirements is generally smooth, with a well-defined
profile whose average slope can be estimated with sufficient precision. On the
other hand, an RC which fails the above criteria has an ill-defined radial
profile and does not faithfully represent the underlying gravitational
potential. In Appendix D we report the relevant photometric and kinematical
quantities (Table 1) alongside their references (Table 2). 
\smallskip

\noindent
$\bullet$ {\it Sample B.} The RCs of PS95 are co-added to form 22 synthetic
curves, arranged by 11 luminosity bins and 11 velocity bins. Since this
procedure averages out most of the observational errors and non-axisymmetric
disturbances present in the individual RCs, only criterion {\it (c)} is applied
and this only to the most severe cases (67 RCs). Instead, we require that RCs be
extended at least out to $0.8\, R_{opt}$ [i.e., a relaxation of criterion {\it
(a)}], which leads to the elimination of 284 RCs. The final set comprises 616 
curves. We derive the synthetic RCs of Sample B as follows.

\noindent
{\it (i)} We normalize each RC to $V_{65}$, the velocity at the radius encompassing 
$65\%$ of the integrated light: for a self-gravitating exponential disc, $V_{65}$ 
corresponds to the radius ($2.2\,R_D$) where the circular velocity peaks (see 
Freeman 1970). With this normalization, at any radius it can immediately be seen 
whether the light traces the dynamical mass, by looking at the {\it amplitude} 
and at the {\it profile} of a given rotation curve. 

\noindent
{\it (ii)} We divide the 616 RCs into 11 luminosity intervals spanning the whole
$I$-band luminosity range $-16.3 < {\rm M}_I< -23.4$, and into 11
velocity-amplitude intervals spanning the range $40 < V_a/($km s$^{-1}) < 340$
($V_a$ is taken from Mathewson et al. 1992). Each interval, of central
luminosity log$\,L$ (velocity amplitude $V_a$), typically contains 70 RCs. Within
each interval, the normalized velocity data are co-added to obtain the
corresponding raw synthetic RCs, $V(R; {\rm log}\,L)$ and $V(R;\,V_a)$. Each of
these has about $\sim 1500$ velocity measurements, and represents the (average)
RC for a spiral of luminosity log$\,L$ (velocity amplitude $V_a$). 

\noindent
{\it (iii)} Finally, we form the smoothed synthetic RCs by averaging the data
in radial bins of size $0.1 \,R_{opt}$. 

In Fig.1 (and Fig.14 in Appendix B) we show the luminosity (velocity) sequence 
of the synthetic RCs. We find that, in the region $0.4 \mincir R/R_{opt} \mincir 
1.1$, the RCs are essentially straight lines, justifying the procedure of Persic \& 
Salucci (1988, 1990b) in deriving their slopes. The present analysis, by co-adding 
several hundred RCs, is negligibly affected by: {\it a)} errors in the assumed 
galaxy distances and/or inclinations; {\it b)} non-axisymmetric disturbances in 
the RCs; and {\it c)} any kind of observational errors. Moreover, we do not assume 
here a specific functional form to represent the RC. As a result, we confirm 
(Persic \& Salucci 1990b) that the {\it inner slope} is a strong (inverse) 
function of luminosity: steeply rising curves and slightly falling curves are 
found at the two opposite extremes of the luminosity range, $\nabla \simeq 0.5$ at 
$M_I=-17$ and $ \nabla \simeq -0.1$ at $M_I=-24$. 

\begin{figure}
\par
\centerline{\psfig{figure=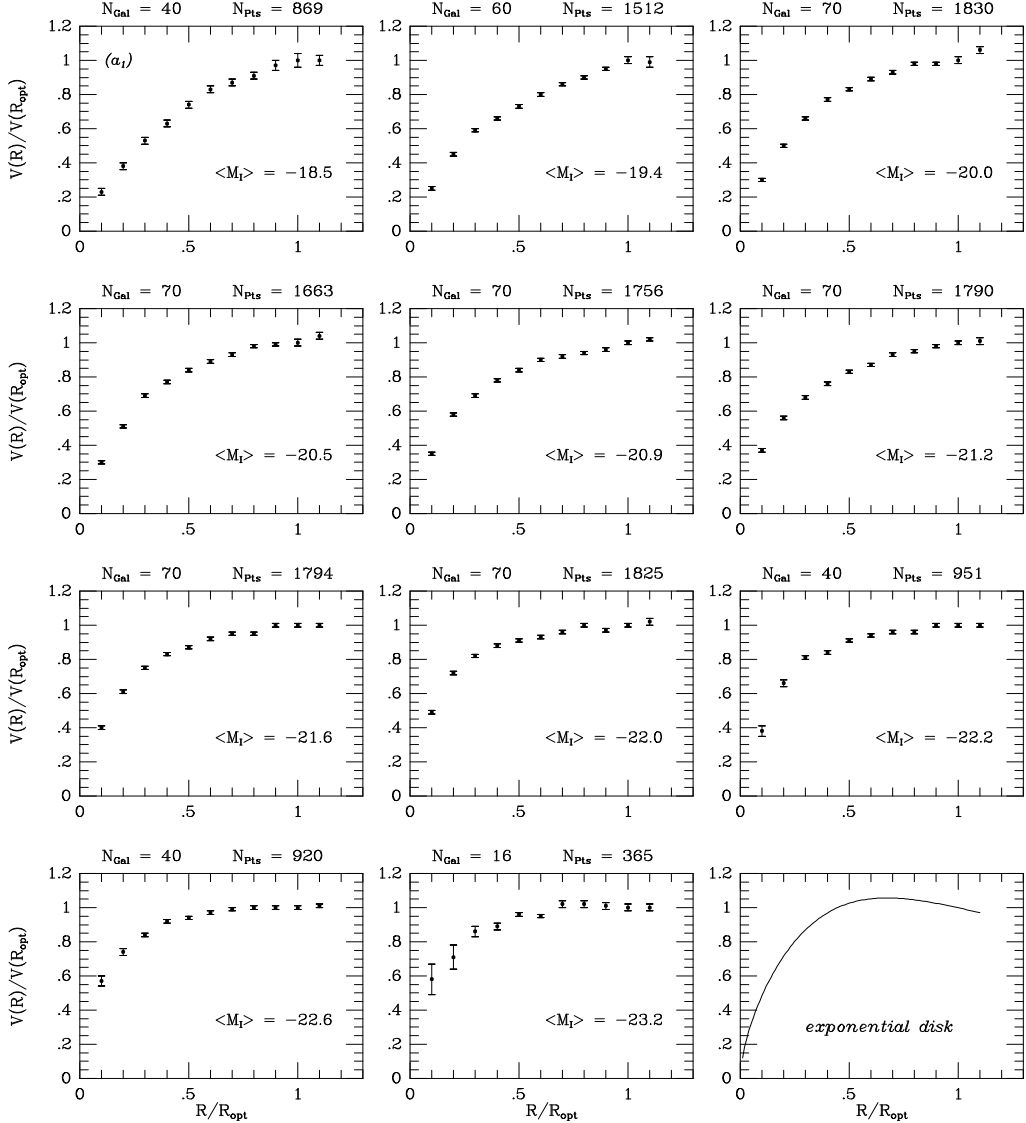,height=15cm,width=15cm}}
\par
\caption{ Synthetic rotation curves for Sample B arranged by luminosity. 
Galactocentric radii are normalized to $R_{opt}$, the radius encompassing
$83\%$ of the total $I$ luminosity. The last panel shows the rotation
curve predicted for a pure self-gravitating exponential thin disc.}
\end{figure}

The co-addition procedure averages over empirical properties of the galaxies such 
as Hubble type, strength of bulge, colour, spiral design, and the presence or 
absence of bars. Since these quantities show, at most, a very loose trend with 
luminosity (de Vaucouleurs et al. 1991; Biviano et al. 1991; Elmegreen \& Elmegreen 
1990), we leave the study of the dependence of the Universal Rotation Curve on 
galaxy morphology to a separate paper (see, however, Section 5).

\section{DM Indicators: Systematics of DM}

The {\it existence} of DM in spiral galaxies is evident from the above 
considerations. The next step, i.e. to derive the mass {\it structure} of the 
spirals, requires assumption of a mass model in order to analyse the rotation 
curves appropriately. However, Persic \& Salucci (1988, 1990b) have shown that the 
main features of the dark matter distribution can be obtained more 
straightforwardly by comparing the light profile and the mass profile (i.e. the RC 
shape): the inability of the luminous matter to trace the distribution of 
gravitating matter becomes the signature of a dark component. In detail, given a 
sample with photometric and kinematical data, at $R_{opt}$ we compute: {\it (a)} 
the slope $\nabla_{lum} \equiv (d\, {\rm log} V_{lum}/ d\, {\rm log} R)_{R_{opt}}$, 
where $V_{lum}(R)$ is determined by the self-gravity of the visible matter derived 
from the surface brightness profile [with no assumption on $(M/L)_{lum}$]; and {\it 
(b)} the actual slope of the rotation curve $\nabla \equiv (d\, {\rm log} V/ d\, 
{\rm log} R)_{R_{opt}}$. $\nabla_{lum}$ and $\nabla$ are computed using the 
average slopes of $V_{lum}(R)$ and $V(R)$ in the range $(0.6 - 1)\, R_{opt}$ (see, 
e.g., Persic \& Salucci 1990b). 

Let us now introduce the quantity $\nabla-\nabla_{lum}$ which compares the 
distribution of gravitating matter with that of the luminous matter, ensuring a
model-independent detection of a mass discrepancy inside $R_{opt}$. In fact,
the light traces the mass as long as $\nabla - \nabla_{lum} \leq 0.04$ (where
the r.h.s. term represents the typical uncertainty in $\nabla$), while a strong
disagreement between the two slopes (i.e., $\nabla - \nabla_{lum} \magcir 0.2$)
implies the presence of a non-luminous component. The difference has an immediate 
interpretation because, given the similarity of the surface-brightness profiles 
of normal spirals (see Fig.12) and the relative smallness of the bulge and gas 
contributions, $\nabla_{lum}$ does not depend on luminosity. In fact we have 
estimated (see Appendix A) that 
$$
\nabla_{lum} ~= ~-0.24 \pm 0.03 \,.
\eqno(2)
$$ 
Therefore the quantity $\nabla+0.25$ probes a mass discrepancy inside $R_{opt}$
that we will interpret in terms of a dark component. The values of $\nabla$ are 
reported in Table 3a for the 131 individual RCs of Sample A, and in Tables 3b,c 
for the 11 + 11 synthetic RCs constructed from Sample B.


\begin{figure}

\par
\centerline{\psfig{figure=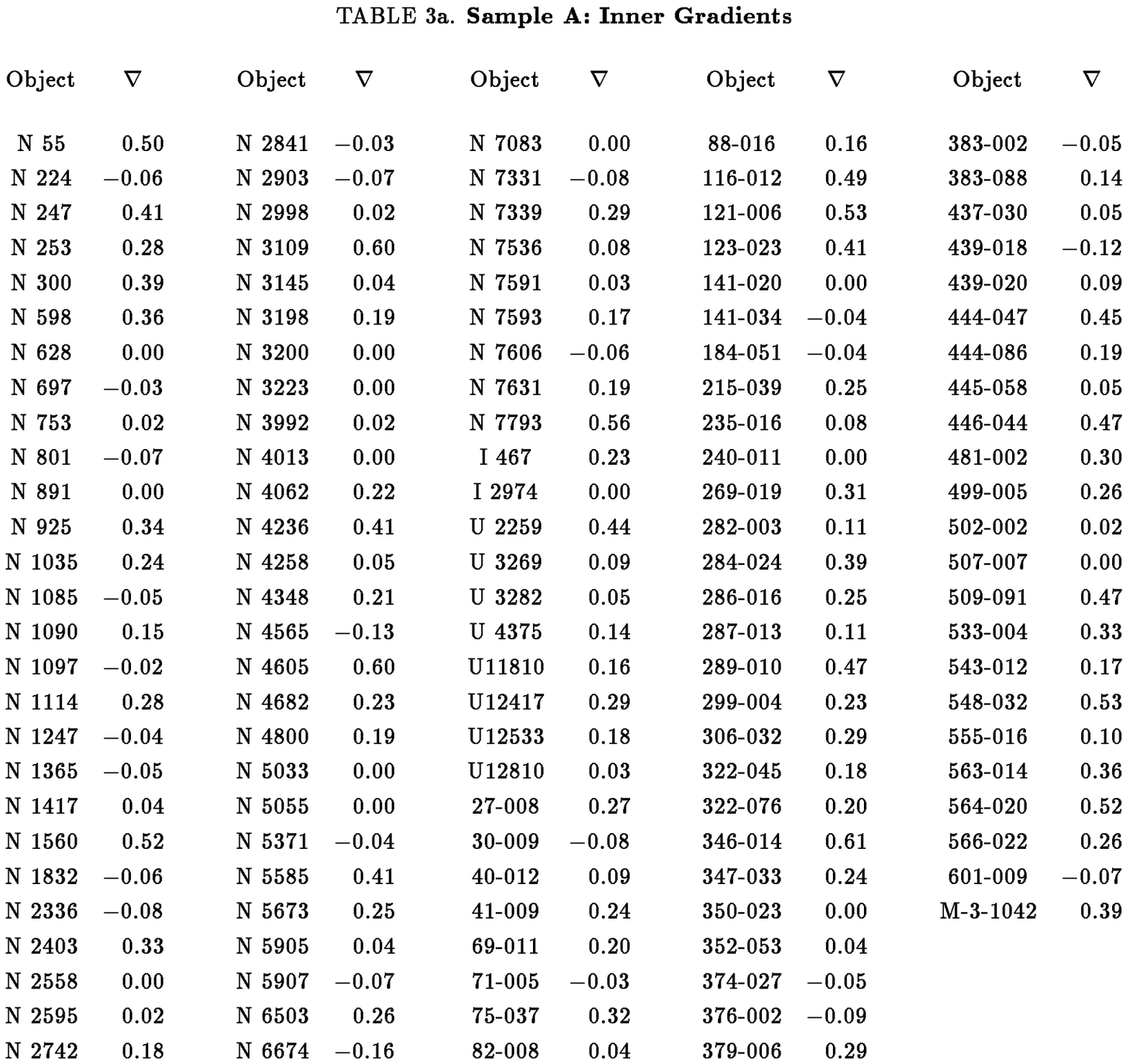,height=11cm,width=12cm}}
\par
\par
\centerline{\hbox{
\psfig{figure=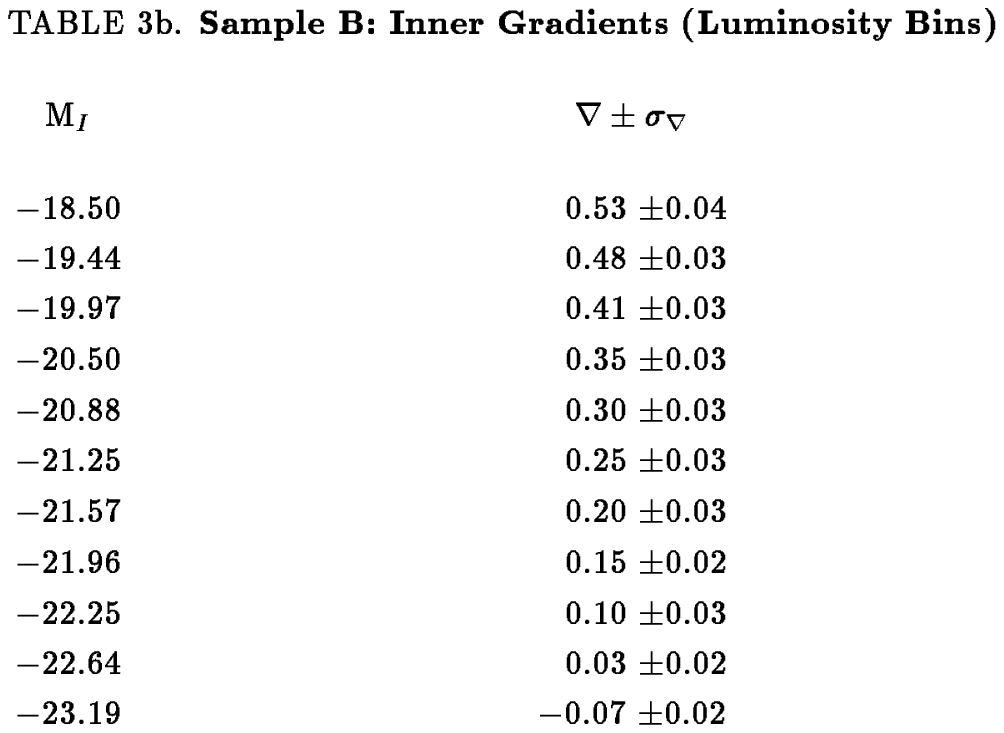,height=5cm,width=7cm}
\psfig{figure=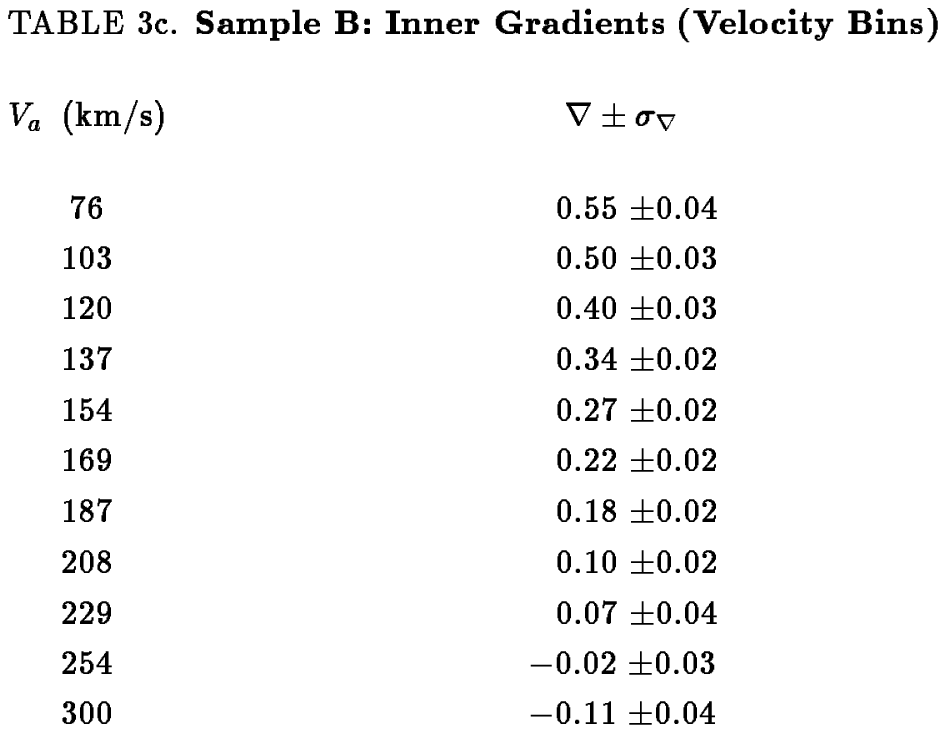,height=5cm,width=7cm}
}}
\par
\end{figure}

 
In our samples many rotation curves are extended beyond the optical size, out to 
$\sim 2 \,R_{opt}$. At these radii the disc and bulge contributions have virtually 
reached the Keplerian regime, $V \propto r^{-1/2}$, and such fall-off must be 
apparent in the RC unless the stellar component is negligible. Let us introduce a 
mass-discrepancy indicator that tests the observed fractional variation in 
velocity, $\delta \equiv [V(2R_{opt})-V(R_{opt})]/V(R_{opt})$, with the variation 
$\delta_{lum}$ that we would expect if the stars and the gas actually traced the 
mass. Consequently, the quantity $\delta - \delta_{lum}$ indicates the level of 
mass discrepancy inside $2 R_{opt}$. The outer gradients $\delta$ are mostly 
estimated from HI data: in Sample A we found a subset of 27 RCs extended to about 
$2\, R_{opt}$ (see Table 1 in Appendix D). In addition, about 200 optical RCs 
of Sample B have measurements farther out than $R_{opt}$: these data, though 
relatively sparse, well supplement the HI data (see Appendix B). Then we can 
build the largest sample for which the haloes' RCs can be traced directly from the 
observed RCs, in that the latter are extended out to radii where the disc 
contribution is certainly negligible. The gradient $\delta_{lum}$ can easily be 
estimated from surface photometry and the HI density distribution: we find that 
this quantity is uncorrelated with galaxy luminosity or velocity amplitude (see 
Appendix A): 
$$
\delta_{lum} ~= ~-0.25 \pm 0.02.
\eqno(3) 
$$
The mass-discrepancy indicator is then simply given by $\delta +{1\over {4}}$.
Values of $\delta \magcir 0$ imply for $R\simeq 2R_{opt}$ a strong decoupling 
between mass and light. 

With these indicators we now examine the presence of dark matter in spirals. 
\bigskip

\subsection{Dark Matter inside $R_{opt}$}
\smallskip

First, we quantify in a simple way the trend between RC profiles and luminosity. 
In Fig.2 we plot $\nabla$ versus luminosity (velocity) for the objects of sample 
A. An inverse proportionality between slope and luminosity (velocity) emerges very 
strongly: the probabilities of a random occurrence are $<10^{-5}\%$. The constant 
of proportionality is $-0.39 \pm 0.02$ ($-1.26 \pm 0.06$), the linear correlation 
coefficient is 0.83 (0.87), and the scatter is $\sigma_{\nabla;\,L_B}= 0.114$ 
($\sigma_{\nabla;\,V(R_{opt})}= 0.099$). Remarkably, the co-added sample shows this 
very same trend but with an even higher level of statistical significance and less 
noise (see Fig 2). This suggests that part of the scatter around the above 
relationships arises from observational errors or kinematical disturbances, 
obviously suppressed by the the co-addition procedure: at a given luminosity 
(velocity amplitude), the intrinsic variance of $\nabla$ is then $\sigma_\nabla 
\mincir 0.05$, one order of magnitude smaller than the range of $\nabla$ spanned 
by spirals. From the combined two samples we obtain, with probability of random 
occurrence $<10^{-6}$, 
$$
\nabla ~ = ~ 0.10 ~- ~0.36 ~ \times ~ {\rm log}{L \over L_*} 
\eqno(4a)
$$
$$
\nabla ~ = ~ 0.10  ~-~ 1.35 ~ \times ~ {\rm log}\, {V_{opt} \over 200}\,,
\eqno(4b)
$$
where the uncertainties on the coefficients are less than 8\%. We remark that the 
inner slope $\nabla$ turns out to be a tight and continuous relationship of 
luminosity, as found in PS91. In disagreement with the claim of Flores et al. 
(1993) (see Section 5), the RCs are {\it very steep} at low luminosities (velocity 
amplitudes), and become {\it gradually flat} --- and eventually {\it slightly 
decreasing} --- at higher luminosities (velocities). 

\begin{figure}
\par
\centerline{\psfig{figure=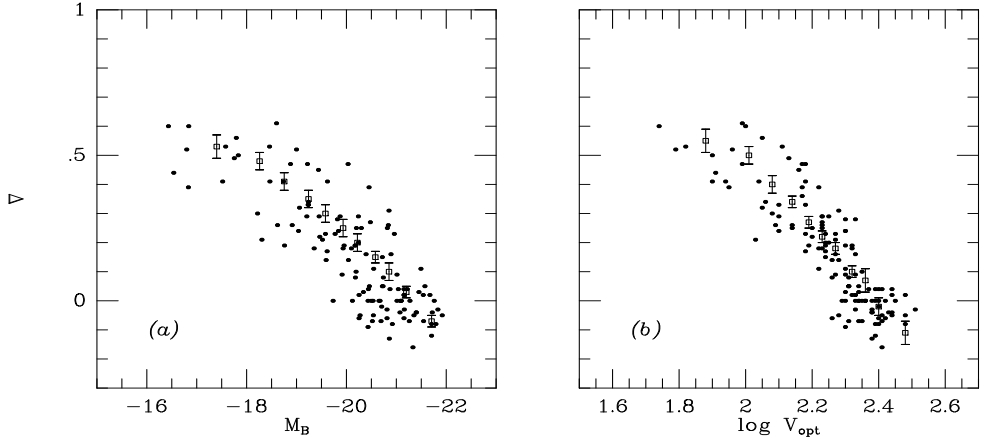,height=7cm,width=15cm,angle=-90}}
\par

\caption { 
The RC slope in the region $(0.6-1)\,R_{opt}$ versus absolute magnitude 
{\it (a)} and versus (log of) rotation amplitude {\it (b)} [$V_{opt} \equiv 
V(R_{opt})$]. The filled circles represent the individual curves of Sample
A; the empty squares represent the synthetic curves of Sample B. The 
conversion between $I$ and $B$ magnitudes has been obtained by
cross-correlating Sample B with the RC3 catalog (de Vaucouleurs et al. 1991): 
M$_B=-0.38+0.92$M$_I$ (129 objects; this relation will be used throughout the
paper).
} 
\label {Figure 3a,b}
\end {figure}


Recalling that $\nabla_{lum} \simeq -0.25$, relationship (4) has two 
straightforward implications: {\it (a)} in every galaxy the LM is unable to 
account for the observed kinematics out to $\sim R_{opt}$; and {\it (b)} 
this inability is marginal (though clearly detectable) at high luminosities, 
and becomes progressively more serious with decreasing luminosity. 

We estimate the minimum mass, $(1-\beta_{max})G^{-1}V^2(R_{opt})R_{opt}$, of the 
dark component residing inside $R_{opt}$ by assuming the {\it maximum-disc 
hypothesis}: 
$$
\beta_{max}~ =~ {3\over 4}\, (1-\nabla) ~\simeq~
	  {2 \over 3} ~+ ~{1 \over 4}~ {\rm log} {L \over L_* }
\eqno(5)
$$
(for details see Persic \& Salucci (1990c). Then, in the biggest (brightest) 
spirals the dark matter comprises at least $\simeq 15 \%$ of the total mass, while 
in the smallest (faintest) ones this fraction increases to $\simeq 85\%$. 
\bigskip
 
\subsection{Dark matter inside $2 R_{opt}$}

In Fig.3 we plot $\delta$ versus luminosity and versus velocity for the very
extended HI RCs of Sample A: a tight outer slope--luminosity (velocity)
relationship emerges (the probability of a chance occurrence is less than
$10^{-4}$). Also shown in Fig.3 are the estimated $\delta$s for the 5 synthetic
extended RCs from Sample B (see Fig.15 in Appendix B): the trend is clearly 
consistent with that of individual RCs from Sample A. All these data are very 
well fitted by: 
$$
\delta ~ = ~ -0.05 ~-~ 0.16 ~{\rm log} {L \over
	      L_* }\,,
\eqno(6)
$$
$$
\delta ~ = ~ -0.05 ~-~ 0.47 ~{\rm log} {V_{opt} \over
	      200 }\,,
\eqno(7)
$$
(with uncertainties of 20\% in zero point and 10\% in slope), which confirms and 
strengthens the results of Salucci \& Frenk (1989) and Casertano \& van Gorkom 
(1991) by means of the consistent evidence of about 100 RCs.

\begin{figure}

\par
\centerline{\psfig{figure=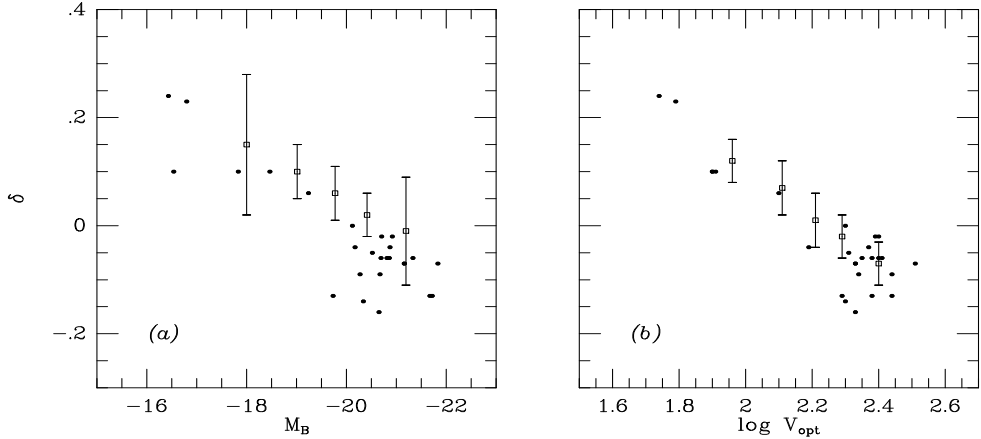,height=7cm,width=15cm,angle=-90}}

\par
\caption {The fractional variation $\delta$ of the rotation velocity 
between $R_{opt}$ and $2\,R_{opt}$ versus absolute magnitude {\it (a)} 
and versus the (log of) rotation velocity at $R_{opt}$ {\it (b)}. 
Filled circles and empty squares denote, respectively, the individual 
objects in Sample A and the extended synthetic RC's from Sample B (given 
in Fig.B2 of Appendix B).}
 
\label{Figure 4a,b}

\end{figure}

By recalling that $\delta_{lum} \simeq -0.25$, we can state that, at $\sim
2\, R_{opt}$, the luminous matter is {\it totally unable} to account for
the observed kinematics.

\subsection{Implications}

The main implications of the above results can be summarized as follows. 

\noindent
$\bullet$ The fraction of DM increases with radius: in fact, if $(1-\beta)$ is 
the fraction of dark matter inside $R_{opt}$, at $2\,R_{opt}$ this fraction has 
increased to $\simeq 1- {\beta \over{ 2(1+\delta)^2}}$. 

\noindent
$\bullet$ The maximum disc mass is $M_{LM}^{max} \sim 2.6 \times 10^{11} h_{75}^
{-1} M_\odot$, while the minimum disc mass is found to be $M_{LM}^{min} \sim 7 
\times 10^{8} h_{75}^{-1} M_\odot$. 

\noindent
$\bullet$ The fraction of dark matter increases, at any normalized radius 
$R/R_{opt}$, with decreasing galaxy luminosity. 

\noindent
$\bullet$ The dark component is much less concentrated than the visible one: in
fact, since $\rho_{DM} \propto V^2/R^2$ and $\rho_{LM} \propto R^{-2} e^{-3.2\,
R/R_{opt}}$, the dark-to-visible density ratio at $2 R_{opt}$ is 15--30 times 
larger than the value at $R_{opt}$. 

Thus the ensuing picture of a spiral galaxy consists of a stellar disc embedded in 
a dark component with a very different density profile. Spirals of high luminosity 
(M$_B <- 21$) have a small dark matter content inside $R_{opt}$ -- consequently, 
their RCs show the {\it kinematical signature}, $\delta_{lum} < \delta < 0$, of the
transition from an inner, LM-dominated region, to an outer, DM-dominated region 
(Salucci \& Frenk 1989). Conversely, low-luminosity (M$_B \magcir -20$) objects 
are dominated nearly everywhere by the DM component, $\delta >0$, so that the 
transition occurs well inside $R_{opt}$ where $V_{lum}$ has still an {\it 
increasing} or {\it flattish} profile not dissimilar to that of the DM, and 
therefore no feature marking the transition between the two regimes appears in 
$V(R)$.

\section{The Universal Rotation Curve}

The luminosity specifies the entire rotation field $V(R, L)$ of spirals. From 
samples A and B we obtain, and plot in Fig.4 the synthetic rotation curve $V(R; 
L)$ [and $V(R; V_a)$] as a function of radius and of galaxy luminosity (velocity). 
These are constructed by combining the co-added rotation curves with the 
slope-velocity-luminosity relationships (6),(7). The rms errors of the 11 $\times$ 
2 co-added RCs are typically $2\pm 1 \%$ inside $R_{opt}$, while beyond $R_{opt}$ 
they increase by up to a factor of 2.

\begin {figure}
\par
\centerline{\psfig{figure=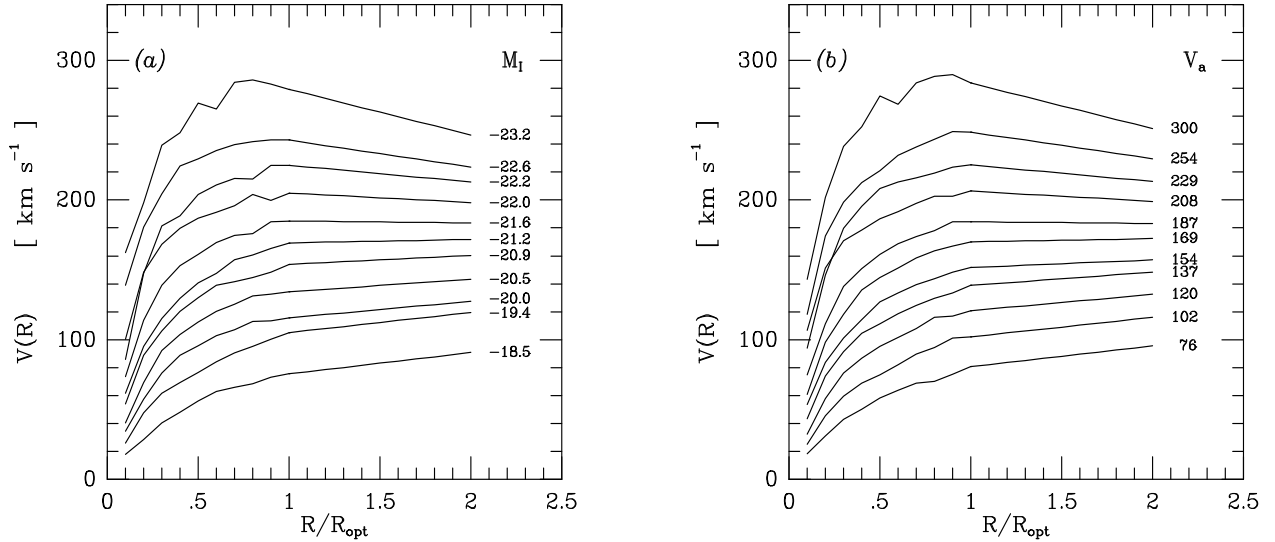,height=10cm,width=18cm}}
\par

\caption {The Universal Rotation Curve of spiral galaxies. Radii are in
units of $R_{opt}$.}

\label {Figure 5} 

\end {figure}

Crucially, $V(R;L)$ has a very small cosmic variance. In fact, the percentage 
residuals $\Delta$ between the individual RCs, $v_i(R;L_i)$ (given in Table 2 of 
PS95), and the averaged RCs, $V(R; L)$, are generally within the estimated 
observational errors ($2-4$ per cent). In Fig.5 we show the comulative distribution 
of $\Delta$: only about $\sim 1/6$ of points has a variance that cannot be simply 
accounted for by observational errors. However, also in these cases, the variance 
is rather small ($\sim 15\%$), and furthermore may arise from a neglected 
dependence of RCs on Hubble Type, or from non-gaussian observational errors, rather 
than indicating actual violations of the Universal Rotation Curve paradigm. Thus, 
as in PS91, we here claim the existence of a {\it Universal Rotation 
Curve}, $V(R;L)$ (given in numerical form by 'Table URC' \footnote{In /pub/psrot at 
147.122.2.158 via anonymous FTP.} and shown in Fig.4). (Sa-type galaxies will be 
investigated in a separate paper.)
\begin{figure}
\par
\centerline{\psfig{figure=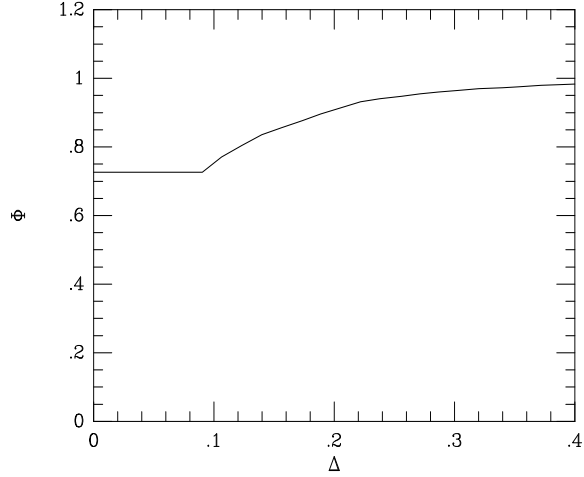,height=7cm,width=9cm}}
\par
\caption {The cumulative distribution of the relative differences between
individual and averaged RCs (all luminosity bins are used).}
\label {Figure 6}
\end{figure}


We now look for an analytical form for $V(R, L)$, that we define as $V_{URC}(R;L)$. 
First, we stress that a very simple function, i.e. two straight lines smoothly 
joining at $R_{opt}$ with luminosity-dependent slopes, describes well the typical 
profile of RCs (see PS91): the inaccuracies are $\mincir 5\%$ (see Persic, Salucci 
\& Stel 1996). Here we proceed a step further by fitting $V(R; L)$ with a function 
that is still simple, but is nevertheless able to:  {\it (a)} describe the 
innermost parts of the RCs; and {\it (b)} pinpoint the separate contributions of 
the dark and luminous components. In detail: 

\noindent
{\it (i)} The contribution from the stellar disc can be written, for $0.04
R_{opt}< R \leq 2R_{opt}$, as (see Appendix C): 
$$
V_d^2(x)~ = ~V^2(R_{opt}) ~\beta~{1.97~ x^{1.22} \over (x^2+0.78^2)^{1.43}}\,,
\eqno(8)
$$
where $x=R/R_{opt}$ and $\beta \equiv [V_d(R_{opt})/V(R_{opt})]^2$. For an
exponential thin disc, $\beta={1.1\, G  M_{LM}\over V^2 (R_{opt})\,R_{opt}}$; 
\medskip

\noindent
{\it (ii)} The contribution from a dark halo can be well represented by:
$$
V_h^2(x)~ = ~V^2(R_{opt}) ~(1-\beta) ~(1+a^2) {x^2\over{(x^2+a^2})}\,,
\eqno(9)
$$
with $a$ being the 'velocity core radius' (measured in units of $R_{opt}$). We 
normalize $V_{URC}$ by setting $V_{URC}(R_{opt})=V(R_{opt};L)$. Therefore, $(1- 
\beta )= {GM_{DM}\over {V^2(R_{opt}) R_{opt}}}$\footnote{In spherical symmetry. 
Otherwise, $M_{DM}$ represents the effective mass.}. 
\medskip

Then we fit $V(R;L)$ with the universal rotation curve $V_{URC}$ given by: 
$$
V_{URC}(R;\, \beta, a)~=~[V_d^2(R)+V^2_h(R)]^{1/2}. 
\eqno(10)
$$
We find that, with
$$
\beta ~= ~0.72~ +~ 0.44\, {\rm log} {L \over L_*}
\eqno(11a)
$$
$$
a~ = ~1.5 \,  \biggl( {L \over L_*} \biggr)^{1/5}\,,
\eqno(11b)
$$
the universal rotation curve $V_{URC}(R;L)$ reproduces the observed rotation
curves $V(R; L)$ to well within their rms errors (see Fig.6). 

\begin{figure}
\par
\centerline{\psfig{figure=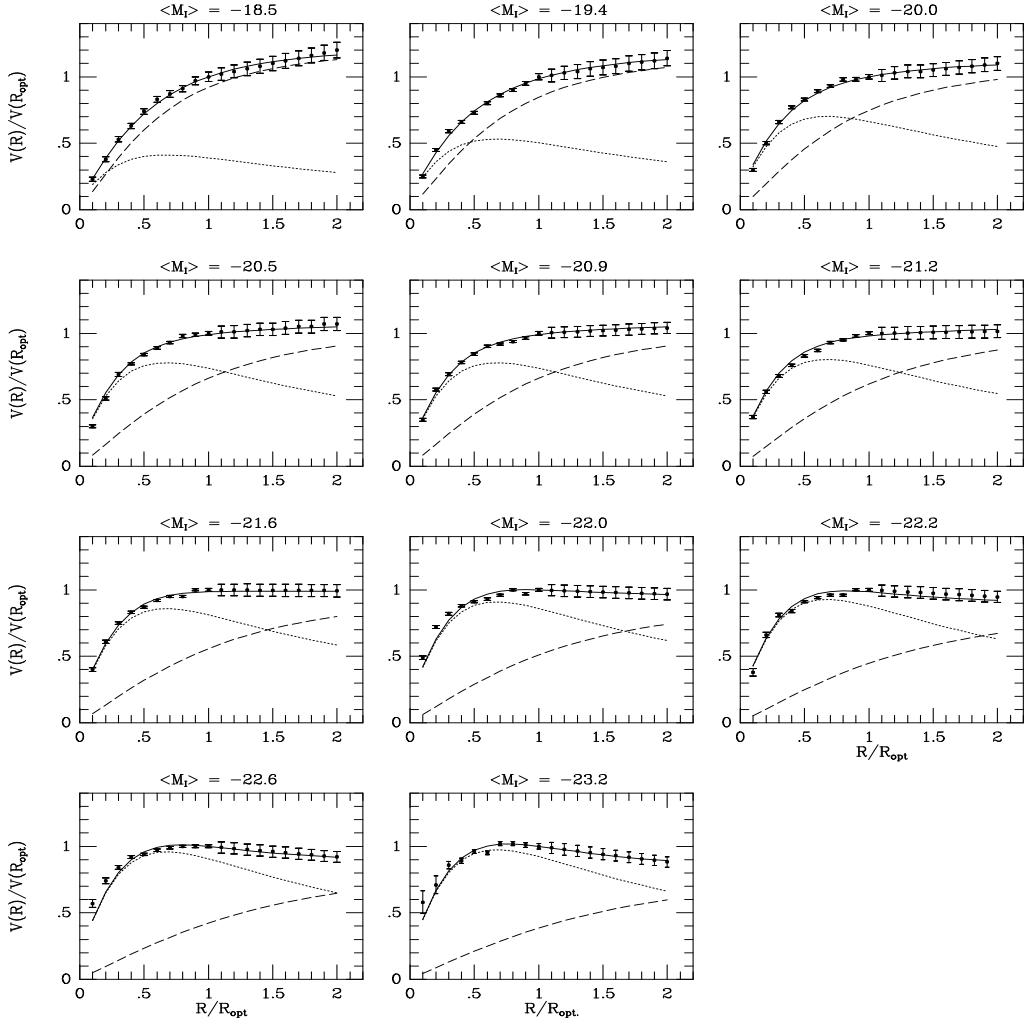,height=15cm,width=15cm}}
\par
\caption {
Best two-component fits to the universal rotation curve (dotted line: disc; 
dashed line: halo). The URC beyond $R_{opt}$ is built by linear extrapolation 
according to eq.(6). Notice that the extent of the RCs and the smallness of 
their rms errors limit the uncertainties on the parameters $\beta$ and $a$ to 
about 10\% and 5\%, respectively.
}
\label{Figure 7} 
\end {figure}
The introduction of a bulge component should further improve the agreement between 
data and fit, by reducing the small discrepancies between eq.(10) and the data for 
very small radii in some synthetic RCs of high luminosity (see Fig.6). In Fig.6 the 
RCs, $V({R \over R_{opt}};L)$, are gauged to the reference frame of the luminous 
matter: the differential effect of the presence of DM emerges clearly as an 
increasing failure of the LM to account for the observed rotation curves as 
luminosity decreases. 

We now set the RCs in the reference frame of the dark matter, whose distribution 
can be characterized by $R_{200}$, i.e. the radius encompassing a mean halo 
overdensity of $<\delta \rho/ \rho>=200$. To smooth the density field we use a 
top-hat filter: $<\rho>_{R_{200}} = M_{DM}(R_{200})/({4 \over 3} \pi R_{200}^3)$, with 
the halo mass $M_{DM}(R)=\int_0^R 4\pi r^2\rho_H(r)\,dr$. (In this formalism the 
'central halo density' is given by lim$_{R \rightarrow 0} {3 V_h^2(R) \over 4 \pi 
G R^2 }$ and does not depend on the local density, $\propto {dM_{DM}/dR \over R^2}$.)
Recalling that the mean mass density of the Universe is $\rho = 3H_0^2 /(8 \pi G)$
\footnote{No result of this paper is changed for $\Omega_0 \neq 1$.}, then 
$R_{200}$ is obtained by solving 
$$
\bigg<{\delta\rho_H \over \rho}\bigg>_{R_{200}} ~=~ 
{2 \over H_0^2} \biggl[ {V^2 - V^2_d \over R^2} \biggr]_{R_{200}}  ~=~ 200\,, 
\eqno(12)
$$
with $V(R)=V_{URC}(R)$ for $R \leq 2\,R_{opt}$, and $V(R)=V_{URC}(2R_{opt})$ 
for $R > 2\,R_{opt}$. The quantity $V^2 - V_d^2 = V_h^2$, appearing in (12), 
can be obtained either directly from (9) and (11) or by using the relation 
$$
V_d^2(R_{200}) ~\simeq ~\biggl( {L \over L_*} \biggr)^{0.4}
{V^2(R_{opt})\, R_{200} \over R_{opt} }
\eqno(13)
$$ 
(see Persic \& Salucci 1990b): the two estimates are in very good mutual
agreement. [In any case, let us remark that $M_{LM} << M_{200}$, with
$M_{200} \equiv M_{DM} (R_{200})$.] In Fig.7 we plot $V({R \over
R_{200}};L)$: we realize that, when scaled to the DM reference frame, the
halo contribution to the RCs are essentially self-similar. Only one
parameter (e.g., the total mass) specifies completely the halo properties
(e.g., the velocity amplitude) far from the region where the LM has
collapsed. 

\begin{figure}
\par
\centerline{\psfig{figure=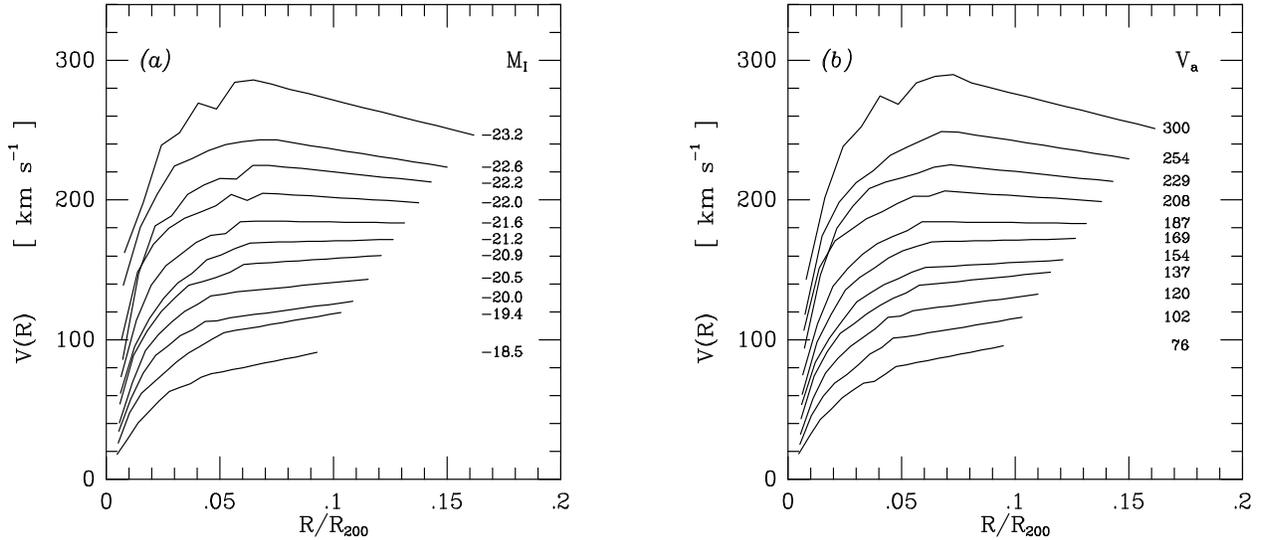,height=10cm,width=18cm}}
\par
\caption{ 
The universal rotation curve of spiral galaxies at different luminosities
and velocities [panels {\it (a)} and {\it (b)}, respectively]. Radii are
in units of $R_{200}$, the radius encompassing a mean halo overdensity of
200, which represents the characteristic scalelength of the DM
distribution. 
}
\label {Figure 8}
\end{figure} 
This evidence (cf. Fig.4) disproves the concept of 'featureless flat RC' and the 
resulting prejudice of a 'cosmic cospiracy', according to which some physical
process would have fine-tuned the structural parameters of the dark and visible 
matter, forcing every RC to be flat and featureless (e.g., Bahcall \& Casertano 
1985). Given the favour often granted to such views (but see Salucci \& Frenk 1989 
and Casertano \& van Gorkom 1991), we want to stress that the large variety of RC 
shapes and the presence of clear features in some of them imply that the physical 
quantities underlying the structure of spirals (e.g., the dark-to-total mass 
ratio) take a wide range of values, though they are inter-related. 

We conclude this section by arguing that, since the stellar mass-to-light ratio in 
spirals spans a range which is limited compared with the range in luminosity, any 
empirical dependence on luminosity may actually reflect a more physical dependence 
on the amount of luminous matter $M_{LM}$.

\section{Discussion}

\noindent
$\bullet$ {\it Where is DM?} It is generally believed that the dark matter 
dominates the outer regions of spirals, while the LM dominates in the innermost
ones. This is only partially true: in low-luminosity galaxies the transition
between the two regimes occurs well inside the optical radius. Let us define the 
transition radius $R_t$ as the innermost radius at which the LM fails to account 
for the observed RC by more than 4 times the typical observational error, i.e.: 
$V_{lum}(R_t) = 0.85 \,V(R_t; L)$. Notice that for $R< R_t$, $V_{lum} \simeq V$. 
So $R_t$ is the innermost radius at which we unambiguously detect the presence of 
dark matter. We also compute the fraction $f_{LM}$ of the total luminous mass 
inside $R_t$: for an exponential thin disc, $f_{LM}= 1-(1+x_t) e^{-x_t}$ with $x_t 
= 3.2R_t/R_{opt}$. In Fig.8 we plot $R_t$ and $f_{LM}$ as a function of luminosity. 
We find that in spirals of low luminosity the mass discrepancy begins before the 
disc half-mass radius $x_t \simeq 1/2$, so that the DM phenomenon is present also 
in the region where most of the ordinary matter resides. Conversely, at high 
luminosities, the phenomenon is evident only where the luminous matter is almost 
absent.  

\begin{figure}

\par
\centerline{\hbox{
\psfig{figure=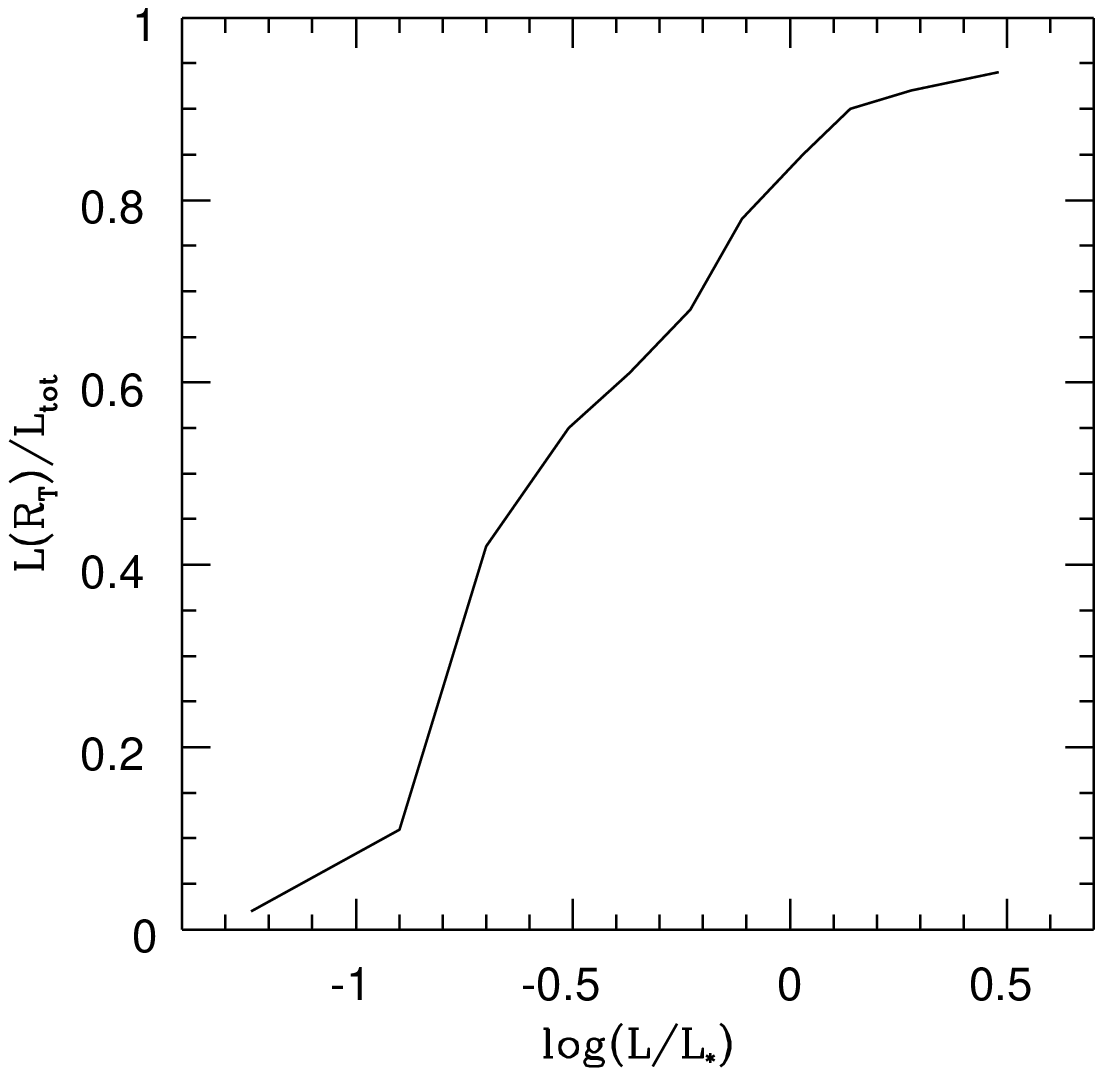,height=7cm,width=8cm}
\psfig{figure=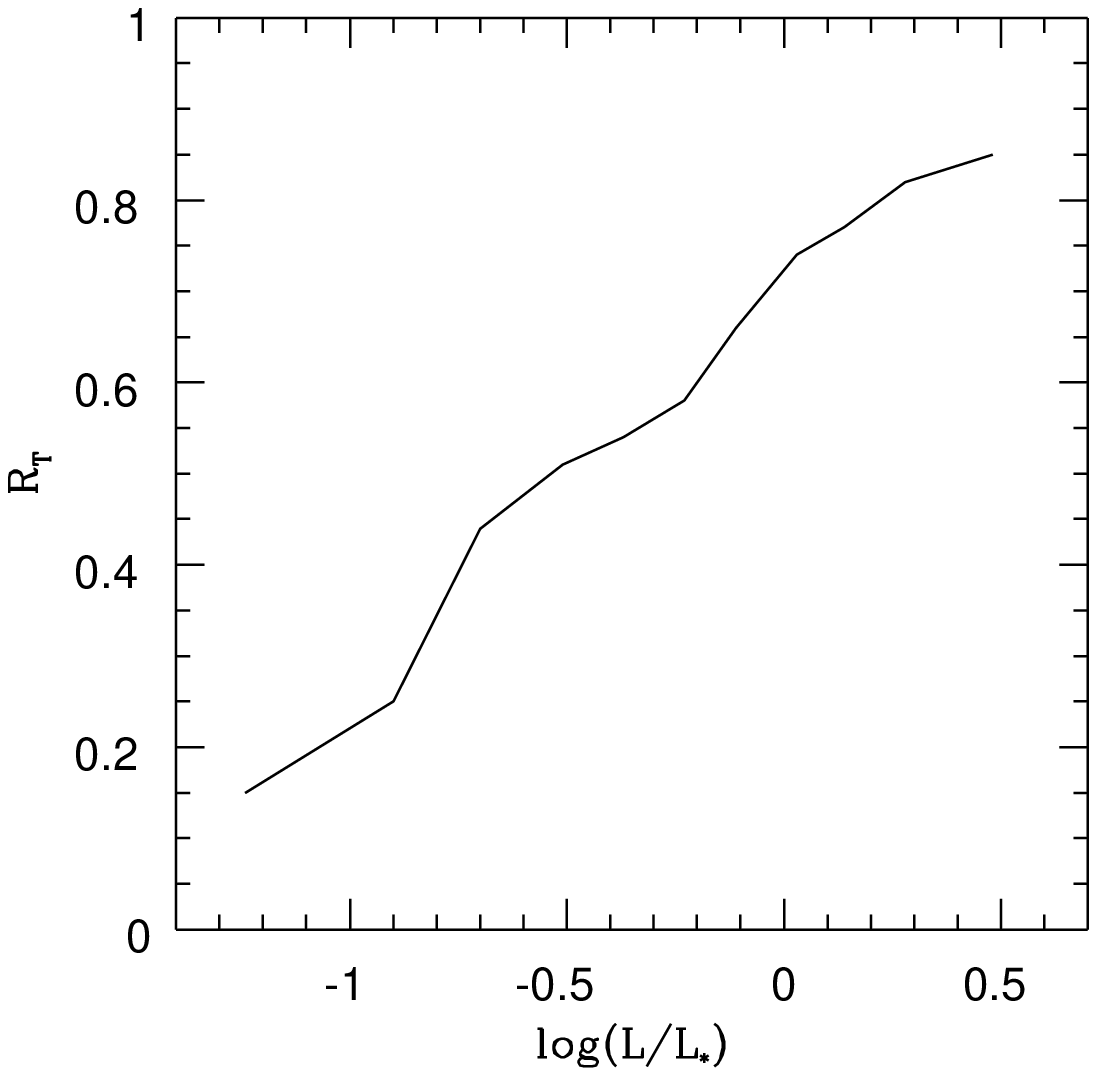,height=7cm,width=8cm}
}}
\caption{
The innermost radius $R_t$ (in units of $R_{opt}$ where the DM is detected
dynamically (left-hand panel) and the fraction of luminous matter (or, 
equivalently, of luminosity) inside $R_t$ (right-hand panel) are plotted 
versus luminosity. ($L_*$ corresponds to M$_B^*=-20.5$ and M$_I^*=-21.9$.)
}
\label{Figure 10}
\end{figure}

\noindent
$\bullet$ {\it The Nature of DM}. We are able to investigate, with a sufficiently 
large sample, the properties of DM in the region where it dominates. In Fig.9 we 
show the {\it local} mass-to-light ratio ${dM(R)\over dR}/ {dL(R)\over dR}$ for 
galaxies of different luminosities. For our purposes we can assume $M(R)= G^{-1}
V^2R$ while the luminosity distribution is obtained from the surface brightness 
profiles. The local $M/L$ in the B-band easily reaches the value of 100, implying 
that the DM cannot be made of low-mass hydrogen-burning stars. Obviously, this 
argument does not rule out other dark baryonic candidates such as black-holes and 
jupiters, but obliges them to {\it (i)} be distributed very differently from the 
luminous baryonic matter, and {\it (ii)} coexist with the luminous matter at about 
$\sim (1/2)\, R_{opt}$. 
\begin{figure}

\par
\centerline{\psfig{figure=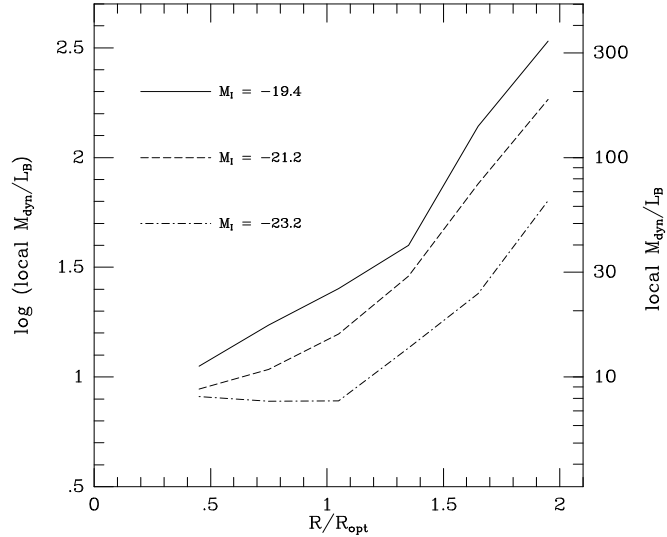,height=9cm,width=8cm}}
\par
\caption{ Local dynamical mass-to-light ratios (in solar units) as a
function of radius at different luminosities. }
\label{Figure 12}
\end{figure}

\noindent
$\bullet$ {\it The Final State of the Baryonic Infall.} Flores et al. (1993) have 
proposed a model for the infall of pre-galactic gas into DM halos which does not 
predict, for $V_{opt} \magcir 100$ km s$^{-1}$, any relationship between the 
luminosity and the RC slope, but rather predicts a random variation around the 
value of $\nabla_{pred}  \simeq 0.1^{+0.1}_{-0.2}$. Flores et al. wondered 
whether the trend between $\nabla$ and luminosity found by PS91 may arise as a 
spurious effect of the small size of the sample used, of the limited extent of 
some RCs and of the assumption of an exponential thin disc taken to derive 
$R_{opt}$. The selection criteria adopted to construct the present samples bypass 
the above-claimed problems: in particular, our data sample is larger by a factor 
15 than that in PS91, and $R_{opt}$ is derived from the actual surface photometry. 
Notwithstanding this (or, rather, just as a result of this), the slope--luminosity 
(velocity) relationship stands out {\it more tightly} than in PS91. We have 
carefully investigated the crucial region $100 \leq V_{opt}/($km s$^{-1}) \leq 
150$, which is very well sampled here: both the individual curves and the 
synthetic curves show a very steep slope ($\nabla >\nabla_{pred} \simeq 0.2$), in 
agreement with the global trend. The latter is very strong, and emerges clearly 
also in samples (as in PS91) with many fewer ($\simeq 50$) objects and lower 
observational accuracy (i.e., $\delta \nabla \simeq 0.1$). Thus, the observed RC 
profiles, unlike the Flores et al. (1993) predictions, are generally very steep, 
and show a marked correlation with luminosity. Navarro et al. (1996) show how an 
infall model {\it \`a la} Flores et al. may naturally reproduce the RC systematics 
highlighted in this (and previous) paper(s).

\section{Conclusion}

In this paper we have investigated the main properties of the mass structure of
spirals. To do this, we have used a very large number ($\sim 1100$) of galaxy RCs, 
to construct: {\it (a)} a sample of 131 high-quality extended RCs; and {\it (b)} a 
sample comprising 616 medium-quality RCs that, co-added in 11 ($\times$ 2) 
synthetic curves, have thouroughly covered the whole luminosity (velocity 
amplitude) sequence of spirals. Our analysis extends out to 2 optical radii and 
spans approximately 6 mag. 

Both samples show that spiral RCs follow a common pattern: their amplitudes and
profiles do not vary freely among galaxies, but depend on luminosity. At low
luminosities the RCs are steep for $R \mincir R_{opt}$, and grow monotonically to
a probably asymptotic value at outer radii. At high luminosities the RCs are flat 
(and even decreasing) for $R \mincir R_{opt}$, and gently fall, from $\sim R_{opt}$ 
outwards, to reach a probably asymptotically constant value farther out. They are 
very well represented by: 
$$
V_{URC}({R\over{\Ro}}) ~=~ V(R_{opt})~ \biggl[ \biggl(0.72+0.44\, {\rm log}
{L \over L_*}\biggr) ~{1.97~x^{1.22}\over{(x^2+0.78^2)^{1.43}}}~+ 
~
1.6\, e^{-0.4(L/L_*)}  { x^2 \over x^2+1.5^2 \,({L\over L_*})^{0.4} } \biggr]^{1/2}
~~~~ {\rm km~s^{-1}}
\eqno(14)
$$
with $x=R/R_{opt}$). The Universal Rotation Curve in eq.(14) (see Fig.10) describes 
any rotation curve at any radius with a very small cosmic variance. In fact eq.(14) 
predicts rotation velocities at any (normalized) radius with a typical accuracy of 
$4\%$. 
\begin{figure}
\par
\centerline{\psfig{figure=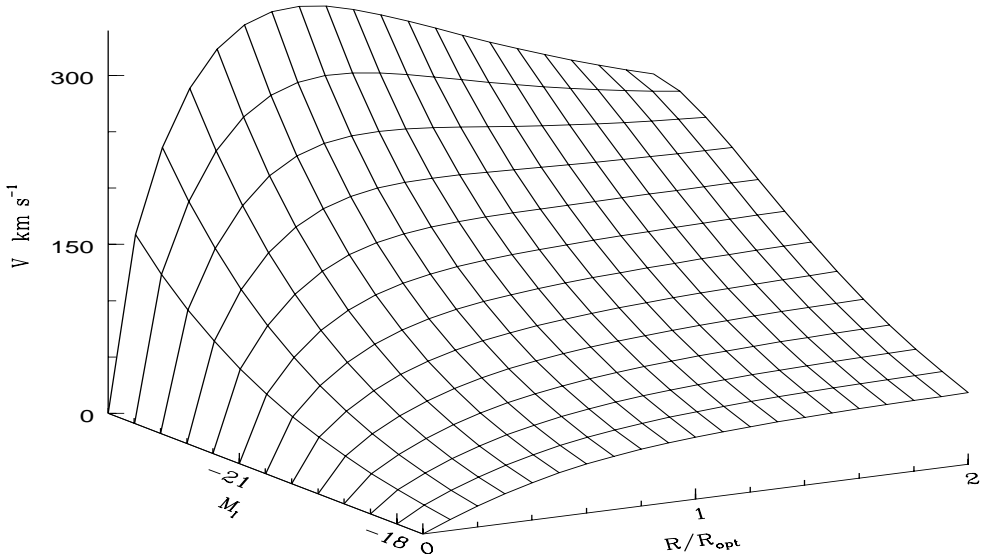,height=10cm,width=15cm}}
\par
\caption{The URC surface}
\label{Figure 12}
\end{figure}

On the other hand, by slicing the URC to match individual observed RCs we can 
derive galaxy luminosities and therefore measure cosmic distances with a typical 
uncertainty of $0.3$ magnitudes. The benefits of using the URC as a distance 
indicator are discussed by Hendry et al. (1996). 

\begin{figure}
\par
\centerline{\psfig{figure=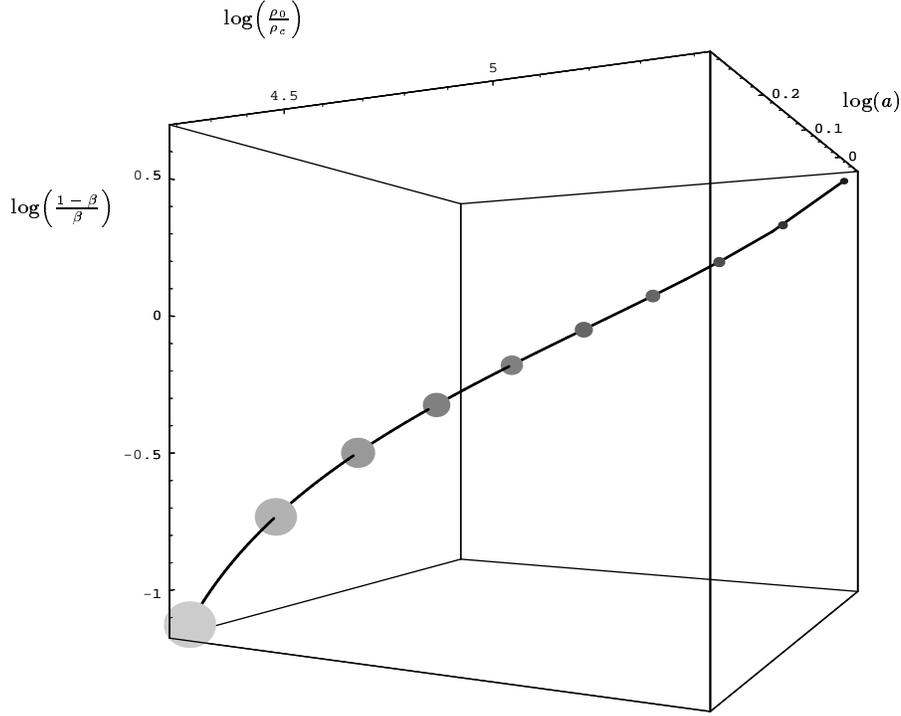,height=12cm,width=12cm}}
\par
\caption{ 
The curve populated by spiral galaxies in the (luminosity)--(dark/visible mass 
ratio)--(DM core radius)--(central DM density) space. (The halo velocity
core radius, $a$, is in units of the optical radius $R_{opt}$.) Different
luminosities are represented as grey circles along the curve with areas
proportional to luminosity at constant $\Delta$log$(L)$ intervals between
$0.06L_*$ and $3L_*$.
} 
\label{Figure 13}
\end{figure}

A particular feature of the Universal Rotation Curve is the strong correlation between the 
shape and the luminosity (velocity) established in previous papers and confirmed 
here over a factor 150 in luminosity (factor of 5 in velocity) and over the variety of the RC 
profiles. This relationship sweeps a narrow locus in the 
profile/amplitude/luminosity space, implying that the great majority of spirals
belong to the same kinematical family. At high luminosity, the profiles of RCs are 
only mildly discrepant from the pure LM predictions, at least within $R_{opt}$, 
so the DM content, although 
detectable, is modest. At low luminosities, on the other hand, the discrepancy between LM predictions 
and RC profiles is very large: the DM is the only relevant mass component. The mass 
structure in spirals is self-similar, but the dark and luminous components are 
coupled with a luminosity-dependent strength. This picture is quantified, by means 
of the results of Sections 3 and 4, by the following relationships (see Salucci et 
al., in preparation):
$$
{\rm log} \biggl( {visible ~mass\over luminosity}\biggr)~ = ~
	0.5 ~+~ 
	0.35\, {\rm log} \biggl( {L \over L_*}\biggr) ~ -~ 
	0.75\, {\rm log}^2 \biggl( {L \over L_*} \biggr) ~~~~~~~~~ 
	{\rm (}B ~ {\rm band; ~ solar ~ units)}  \,,
$$
$$
disc ~central~surface ~density~ = ~
	  780~ \biggl(1~+~0.6 \, {\rm log}{L\over L_*}\biggr)^2 M_\odot 
        {\rm pc}^{-2} ~=~ 
	  235~ \biggl(1~+~{5 \over 3} \, {\rm log}{M_{LM}\over 10^{10} M\odot}
	  \biggr)^2 M_\odot {\rm pc}^{-2}\,,
$$
$$
{dark ~ mass\over {visible ~ mass}} ~= 
	~0.4~ \biggl({L \over L_*}\biggr)^{-0.9} x^3 {1+1.5^2 
	({L\over L_*})^{0.4} \over x^2+1.5^2 ({L\over L_*})^{0.4} } ~=
		~0.16~ \biggl({M_{LM} \over M_{LM}^{max}}\biggr)^{-0.72} 
		x^3 {1+3.4 ({M_{LM} \over M_{LM}^{max}})^{0.32} \over 
		x^2+3.4 ({M_{LM} \over M_{LM}^{max}})^{0.32} }\,,
$$
$$
{halo ~core ~radius \over optical ~ radius} ~= 
		~ 1.5 ~\biggl( {L \over L_*} \biggr)^{0.2} ~=
		~ 1.3 ~\biggl( {M_{200} \over 10^{12}M_\odot} \biggr)^{0.4} \,,
$$
$$
{halo ~core ~radius \over virial ~ radius} ~= 
		~ 0.1 ~\biggl( {L \over L_*} \biggr)^{0.34} ~=
		~ 0.075 ~\biggl( {M_{200} \over 10^{12}M_\odot} \biggr)^{0.62} \,,
$$
$$
{halo~central~density \over critical~density} ~=
		~ 3.5 \times 10^4~ \biggl( {L\over L_*} \biggr)^{-0.7} ~=
		~ 6.3 \times 10^4 ~\biggl( {M_{200} \over 
				10^{12}M_\odot} \biggr)^{-1.3} \,,
$$ 
$$
{virial~radius \over optical~radius}\,=\,
            ~ 14.8~\Bigl({L\over L_*}\Bigr)^{-0.14} ~=
		~ 16.7 ~\biggl( {M_{200} \over 10^{12}M_\odot} \biggr)^{-0.25} \,,
$$
$$
halo ~mass ~=~ 1.6 \times 10^{12} \biggl( {L\over L_*} \biggr)^{0.56} M_\odot ~=
		~ 3 \times 10^{12} \Bigl( {M_{LM} \over M_{LM}^{max}} \Bigr)^{0.44}
              M_\odot\,,
$$
$$
fraction ~ of ~ primordial ~ baryons ~ turned ~ into ~ stars ~=~ 65 
		\Bigl({M_{LM} \over M_{LM}^{max}}\Bigr)^{0.56}
		\Bigl({\Omega_{BBN} \over 0.06\, h_{50}^{-2}}\Bigr)^{-1} \%
$$
(where $\Omega_{BBN}$ is the cosmological density parameter of Big Bang 
synthesized baryons). 

A curve in the (luminosity)--(dark-to-luminous mass ratio)--(DM core 
radius)--(central DM density) 4-D space (see Fig.11) is the geometrical analogue 
of these structural relationships. This curve, which is the counterpart, for 
spirals, of the {\it Fundamental Plane} for ellipticals, shows that the 
distribution of DM is completely different from that of the luminous matter and 
strongly varies with luminosity, and that, especially at low luminosity, the 
presence of DM is already detected well inside the optical regions. The 
implications are as follows:
\medskip

\noindent
$\bullet$ the Universe becomes darker at smaller scales; 

\noindent
$\bullet$ smaller discs have lower densities and mass-to-light ratios;

\noindent
$\bullet$ smaller discs have lost a larger fraction of their original baryon 
content; 

\noindent
$\bullet$ smaller haloes have higher central densities;

\noindent
$\bullet$ the formation of spirals is controlled by just one quantity, e.g. the 
total mass (or, alternatively, the initial DM overdensity).

\bigskip

{\it Acknowledgements.} We thank John Miller for carefully reading the manuscript, 
and the referee for insightful comments that have improved the presentation of this 
work. We also thank Joel Primack for contributing, through several stimulating 
comments on PS91, to our motivation to construct the 967 rotation curve sample 
(PS95) and to undertake this analysis.

\vglue 0.4truecm

\centerline {\bf References }

\vglue 0.1truecm

\ref{Aaronson, M. et al. 1982, ApJS, 50, 241}
\ref{Amram, P., Le Coarer, E., Marcelin, M., Balkowski, C., Sullivan, W.T., III,
    \& Cayatte, V. 1992, A\&AS, 94, 175}
\ref{Amram, P., Marcelin, M., Balkowski, C., Cayatte, V., Sullivan, W.T., III,
    \& Le Coarer, E., 1994, A\&AS, 103, 5}
\ref{Ashman, K.M. 1992, PASP, 104, 1109}
\ref{Bahcall, J.N., \& Casertano, S. 1985, ApJ, 293, L7}
\ref{Begeman, K. 1987, Ph.D. thesis, Groningen University}
\ref{Biviano, A., Girardi, M., Giuricin, G., Mardirossian, F., \& Mezzetti, M.
    1991, ApJ, 376, 458}
\ref{Blackman, C.P., \& van Moorsel, G.A. 1984, MNRAS, 208, 91}
\ref{Boroson, T. 1981, ApJS, 46, 177}
\ref{Bosma, A. 1981a, AJ, 86, 1791}
\ref{Bosma, A. 1981b, AJ, 86, 1825}
\ref{Bottema, R. 1989, A\&A, 221, 236}
\ref{Bottema, R., Shostak, G.S., \& van der Kruit, P.C. 1987, Nature, 328, 401}
\ref{Broeils, A.H. 1992a, A\&A, 256, 19}
\ref{Broeils, A.H. 1992b, Ph.D. thesis, Groningen University}
\ref{Broeils, A.H., \& Knapen, J.H. 1991, A\&AS, 91, 469}
\ref{Burstein, D., \& Heiles, C. 1984, ApJS, 54, 33} 
\ref{Burstein, D., \& Rubin, V.C. 1985, ApJ, 297, 423}
\ref{Carignan, C. \& Puche, D. 1990a, AJ, 100, 394}
\ref{Carignan, C. \& Puche, D. 1990b, AJ, 100, 641}
\ref{Carignan, C., Sancisi, R., \& van Albada, T.S. 1988, AJ, 95, 37}
\ref{Casertano, S., \& van Gorkom, J.H. 1991, AJ, 101, 1231}
\ref{C\^ot\'e, S., Carignan, C., \& Sancisi, R. 1991, AJ, 102, 904}
\ref{de Vaucouleurs, G., de Vaucouleurs, A., Corwin, H.G., Jr., Buta,
    R.J., Paturel, G., \& Fouqu\'e, P. 1991, The Third Reference 
    Catalog of Bright Galaxies (Austin: Univ. of Texas Press) (RC3)}
\ref{Elmegreen, B.G., \& Elmegreen D.M. 1985, ApJ, 288, 438}
\ref{Elmegreen, D.M., \& Elmegreen B.G. 1990, ApJ, 364, 412}
\ref{Evrard, A.E., Summers, F.J., \& Davis, M. 1994, ApJ, 422, 11}
\ref{Faber, S.M., \& Gallagher, J.S. 1979, Ann. Rev. Astr. Astrophys., 17, 135}
\ref{Flores, R., Primack, J.R., Blumenthal, G.R., \& Faber, S.M. 1993, 
    ApJ, 412, 443}
\ref{Freeman, K.C. 1970, ApJ, 160, 811}
\ref{Gottesman, S.T. 1980, AJ, 85, 824}
\ref{Gottesman, S.T., Ball, R., Hunter, J.H., \& Huntley, J.M. 1984,
    ApJ, 286, 471}
\ref{Hendry, M., Rauzy, S., Salucci, P., \& Persic, P. 1996, in 
     Giuricin, G., Mardirossian, F., Mezzetti, M., eds., 
     From Galaxies to Galaxy Systems. 
     Ap.Lett.\&Comm., Gordon \& Breach, New York}
\ref{Hodge, P.W. 1978, ApJS, 37, 429}
\ref{Jobin, M., \& Carignan, C. 1990, AJ, 100, 648}
\ref{Kamphuis, J., \& Briggs, F. 1992, A\&A, 253, 335}
\ref{Kent, S.M. 1985, ApJS, 59, 115}
\ref{Kent, S.M. 1986, AJ, 91, 1301}
\ref{Kent, S.M. 1987, AJ, 93, 816}
\ref{Kent, S.M. 1989, PASP, 101, 489}
\ref{Lake, G., \& Feinswog, L., 1989, AJ, 98, 166}
\ref{Marcelin, M., Boulesteix, J., \& Court\`es, G. 1982, A\&A, 108, 134}
\ref{Mathewson, D.S., Ford, V.L., \& Buchhorn, M. 1992, ApJS, 81, 413}
\ref{Moore, B. 1994, Nature, 370, 629}
\ref{Navarro, J.F., \& White, S.D.M. 1994, MNRAS, 267, 401}
\ref{Navarro, J.F., Frenk, C.S., \& White, S.D.M. 1996, ApJ, 462, 563}
\ref{Newton, K. 1980, MNRAS, 190, 689}
\ref{Ondrechen, M.P., \& van der Hulst, J.M. 1989, ApJ, 342, 29}
\ref{Ondrechen, M.P., van der Hulst, J.M., \& Hummel, E. 1989, ApJ, 342, 39}
\ref{Persic, M., \& Salucci, P. 1988, MNRAS, 234, 131}
\ref{Persic, M., \& Salucci, P. 1990a, ApJ, 355, 44}
\ref{Persic, M., \& Salucci, P. 1990b, MNRAS, 245, 577}
\ref{Persic, M., \& Salucci, P. 1990c, MNRAS, 247, 349}
\ref{Persic, M., \& Salucci, P. 1991, ApJ, 368, 60 (PS91)}
\ref{Persic, M., \& Salucci, P. 1992, MNRAS, 258, 14P}
\ref{Persic, M., \& Salucci, P. 1995, ApJS, 99, 501 (PS95}
\ref{Persic, M., Salucci, P., \& Ashman, K.M. 1993, A\&A, 279, 343}
\ref{Persic, M., Salucci, P., \& Stel, F. 1996, Ap.Lett.\&Comm., 33, 205}
\ref{Puche, D., Carignan, C., \& Bosma, A. 1990, AJ, 100, 1468}
\ref{Puche, D., Carignan, C., \& van Gorkom, J.H. 1991, AJ, 101, 456}
\ref{Puche, D., Carignan, C., \& Wainscoat, R.J. 1991, AJ, 101, 447}
\ref{Regan, M.W., \& Vogel, S.N. 1994, ApJ, 434, 536}
\ref{Rogstad, D.H., Wright, M.C.H. \& Lockhart, I.A. 1976, ApJ, 204, 703}
\ref{Rubin, V.C., Ford, W.K., Jr., \& Thonnard, N. 1980, ApJ, 238, 471}
\ref{Rubin, V.C., Ford, W.K., Jr., Thonnard, N. \& Burstein, D., 1982,
    ApJ, 261, 439}
\ref{Rubin, V.C., Burstein, D., Ford, W.K., Jr., \& Thonnard, N. 1985,
    ApJ, 289, 81}
\ref{Rubin, V.C., Whitmore, B.C., \& Ford, W.K., Jr. 1988, ApJ, 333, 522}
\ref{Salucci, P., \& Frenk, C.S. 1989, MNRAS, 237, 247}
\ref{Sancisi, R., \& Allen, R.J. 1979, A\&A, 74, 73}
\ref{Sancisi, R., \& van Albada, T.S. 1987, in Dark Matter in the Universe
    (IAU Symposium No. 117, ed. J. Kormendy \& G.R. Knapp; Dordrecht: Reidel),
    p. 67}
\ref{Schombert, J.M., \& Bothun, G.D. 1987, AJ, 93, 60}
\ref{Schommer, R.A., Bothun, G.D., Williams, T.B., \& Mould, J.R. 1993, AJ,
    105, 97} 
\ref{Shaw, M.A., \& Gilmore, G. 1989, MNRAS, 237, 903}
\ref{Shostak, G.S. 1973, A\&A, 24, 411}
\ref{Shostak, G.S., \& van der Kruit, P.C. 1984, A\&A, 132, 20}
\ref{Tully, R.B. 1988, Nearby Galaxies Catalog (Cambridge: Cambridge
    University Press)}
\ref{van Albada, G.D. 1980, A\&A, 90, 123}
\ref{van der Kruit, P.C., \& Searle, L. 1981a, A\&A, 95, 105}
\ref{van der Kruit, P.C., \& Searle, L. 1981b, A\&A, 95, 116}
\ref{van der Kruit, P.C., \& Searle, L. 1982, A\&A, 110, 61}
\ref{van Moorsel, G.A. 1982, A\&A, 107, 66}
\ref{van Moorsel, G.A. 1983a, A\&AS, 53, 271}
\ref{van Moorsel, G.A. 1983b, A\&AS, 54, 19}
\ref{Wevers, B.M.H.R., van der Kruit, P.C., \& Allen, R.J. 1986, A\&AS, 66, 505}
\ref{Whitmore, B.C. 1984, ApJ, 278, 71}
\ref{Whitmore, B.C., Forbes, D., \& Rubin, V.C. 1988, ApJ, 333, 542}

\vspace{2cm}

\centerline{ \bf APPENDIX A}

In this Appendix we study the connection between the velocity gradients $\nabla$ 
and $\delta$ and the distribution of DM. We start from the condition of centrifugal 
equilibrium: 
$$
V^2(R) ~=~ V^2_{lum}(R)~+~V^2_h(R)\,,
\eqno(A1)
$$
where $V^2_{lum}(R)$ is the quadratic sum of the three luminous components, i.e.
gas, disc, and bulge. The gas contribution is obtained from the HI surface density, 
and the functional form of the stellar contribution is determined by the surface 
brightness profile (the effect of the bulge is studied in Persic, Salucci \& 
Ashman 1993). 
\begin{figure}

\par
\centerline{\psfig{figure=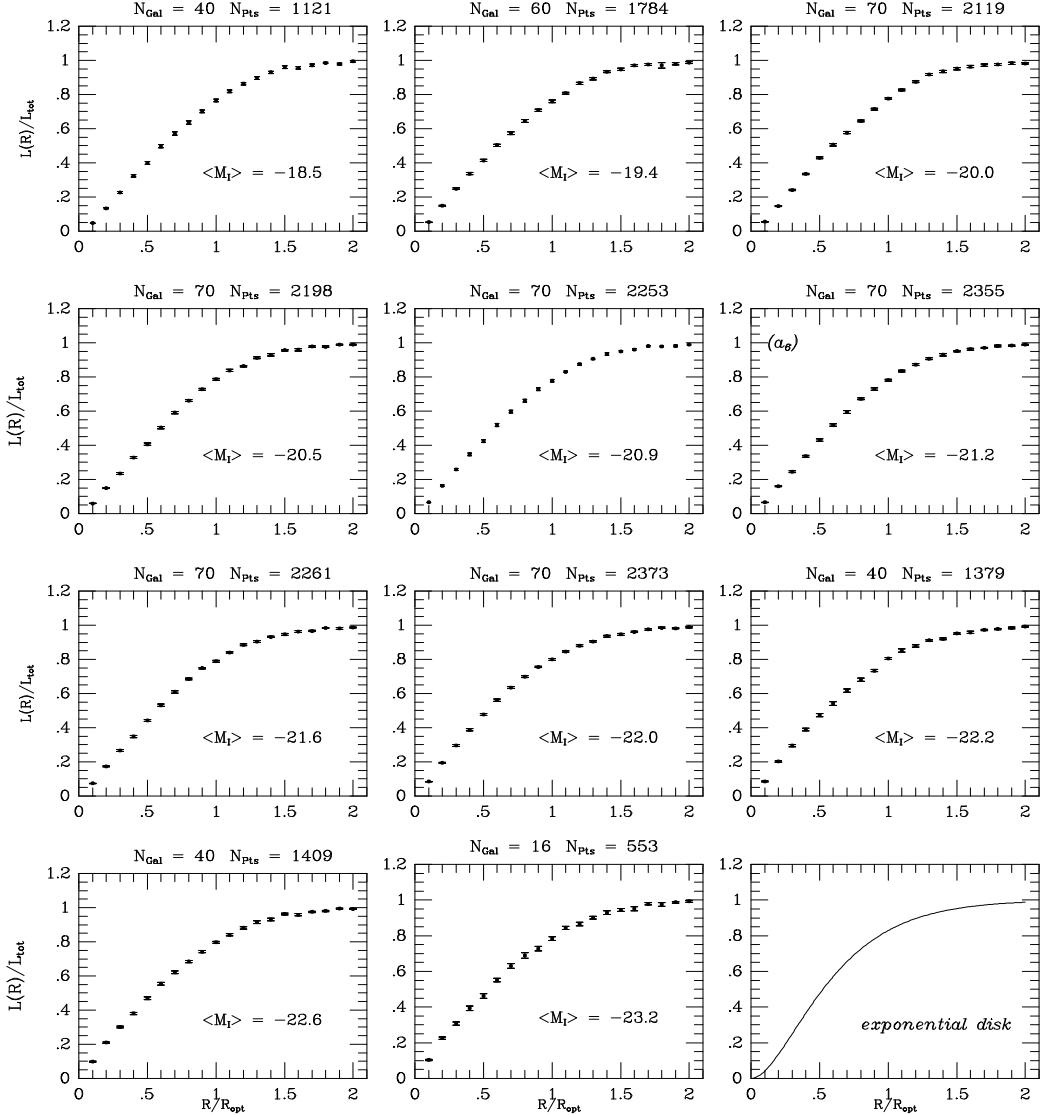,height=15cm,width=15cm}}
\par
\caption{ 
The luminous-mass profiles for the galaxies in Sample B, grouped by
luminosity bins. For each individual object, the light profile $L(r)
\propto \int_0^r I(r^\prime)\,r^\prime \,dr^ \prime$ is normalized to 
its total value $L_\infty$; the radius is normalized to $R_{opt}$.
Grouping the light profiles by velocity amplitude yields a similar 
result.
} 
\label{Figure A1}
\end{figure}
Let us define
$$
\delta_{lum}~= ~{ V_{lum}\, (2R_{opt})- V_{lum} (R_{opt}) 
		\over V_{lum}(R_{opt})}\,, 
\eqno(A2)
$$ 
and
$$
\nabla_{lum} ~= ~{d{\rm log} V_{lum}(R) \over d {\rm log} R}
			\biggr|_{R_{opt}}\,;
\eqno(A3)
$$ 
and define $\delta_h$ and $\nabla_h$ similarly for the halo component. 
Finally, let us define the LM fraction inside the optical radius as: 
$$
\beta \equiv { V^2_{lum}(R) \over V^2(R) } \biggr|_{R_{opt}}\,.
\eqno(A4)
$$
The RC-profile equation [cf. eq.(4) in Persic \& Salucci 1990b] is:
$$
\nabla ~=~ \beta \, \nabla_{lum} ~+~ (1-\beta) \,\nabla_h\,,
\eqno(A5)
$$
while from the definition of $\delta$ we have:
$$
(1+\delta)^2 ~=~ \beta \,(1+\delta_{lum})^2~+~ (1-\beta)\, (1+  \delta_h)^2\,. 
\eqno(A6)
$$
We investigate whether the light profile of spirals depends on luminosity by
co-adding the surface brightnesses of the 616 galaxies of Sample B to form
synthetic light profiles, arranged by luminosity (velocity) bins (see Fig.12).

We stress that the actual light distributions do not depend on luminosity
(velocity), which implies that the velocity profile of the stellar component,
$$
\biggl[ {V_b^2(R) + V_d^2(R) \over V_b^2(R_{opt}) + V_d^2(R_{opt})}
\biggr]^{1/2}\,,
$$
does not depend on luminosity. In particular, at $2R_{opt}$ the
light has already converged: this implies that the bulge and the disc are
already very close to a Keplerian regime, $V_b(r)\propto r^{-1/2}$ and $V_d
\propto r^{-1/2}$, in any object. 

\begin{figure}
\par
\centerline{\hbox{
\psfig{figure=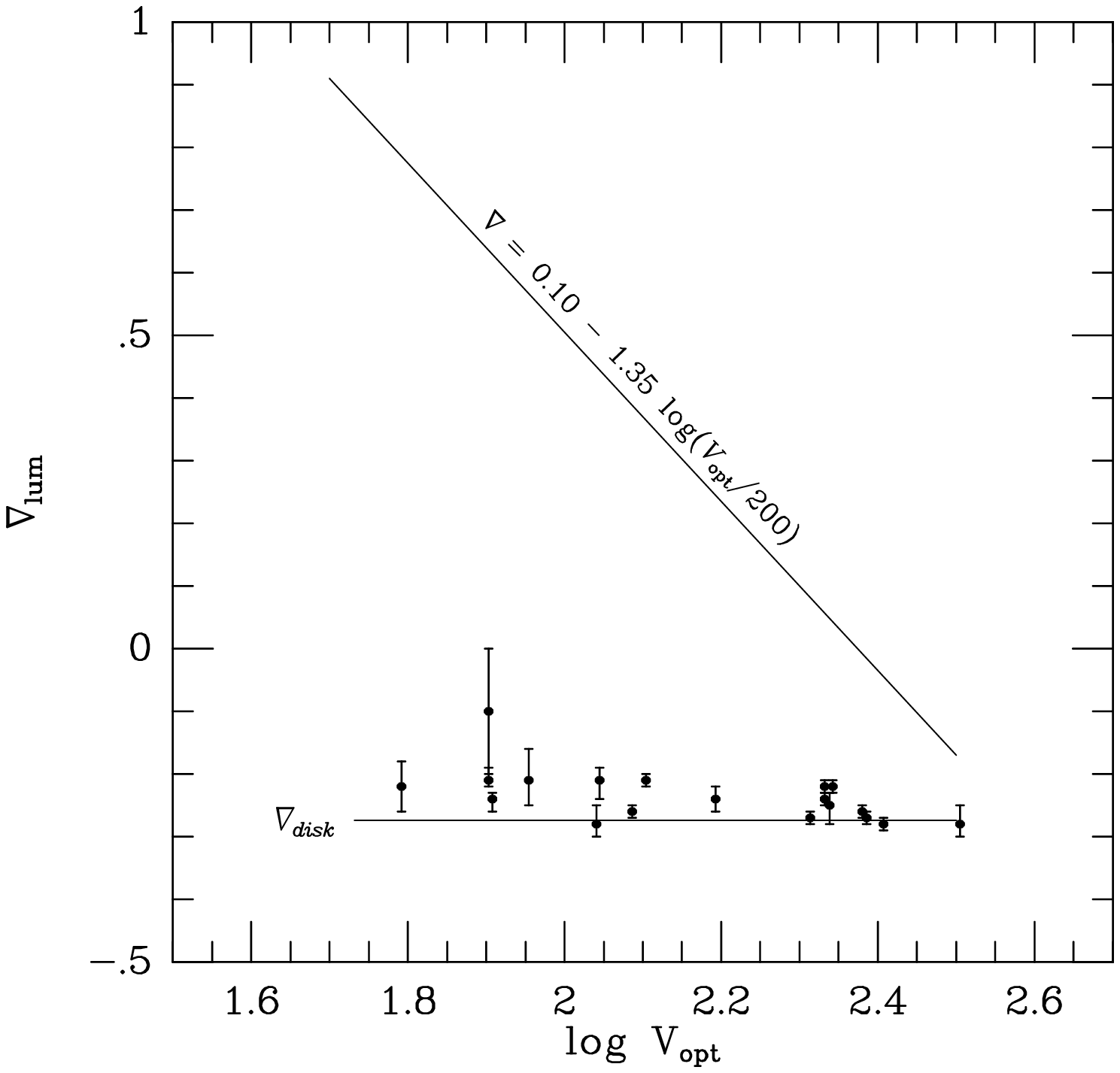,height=8cm,width=8cm}
\psfig{figure=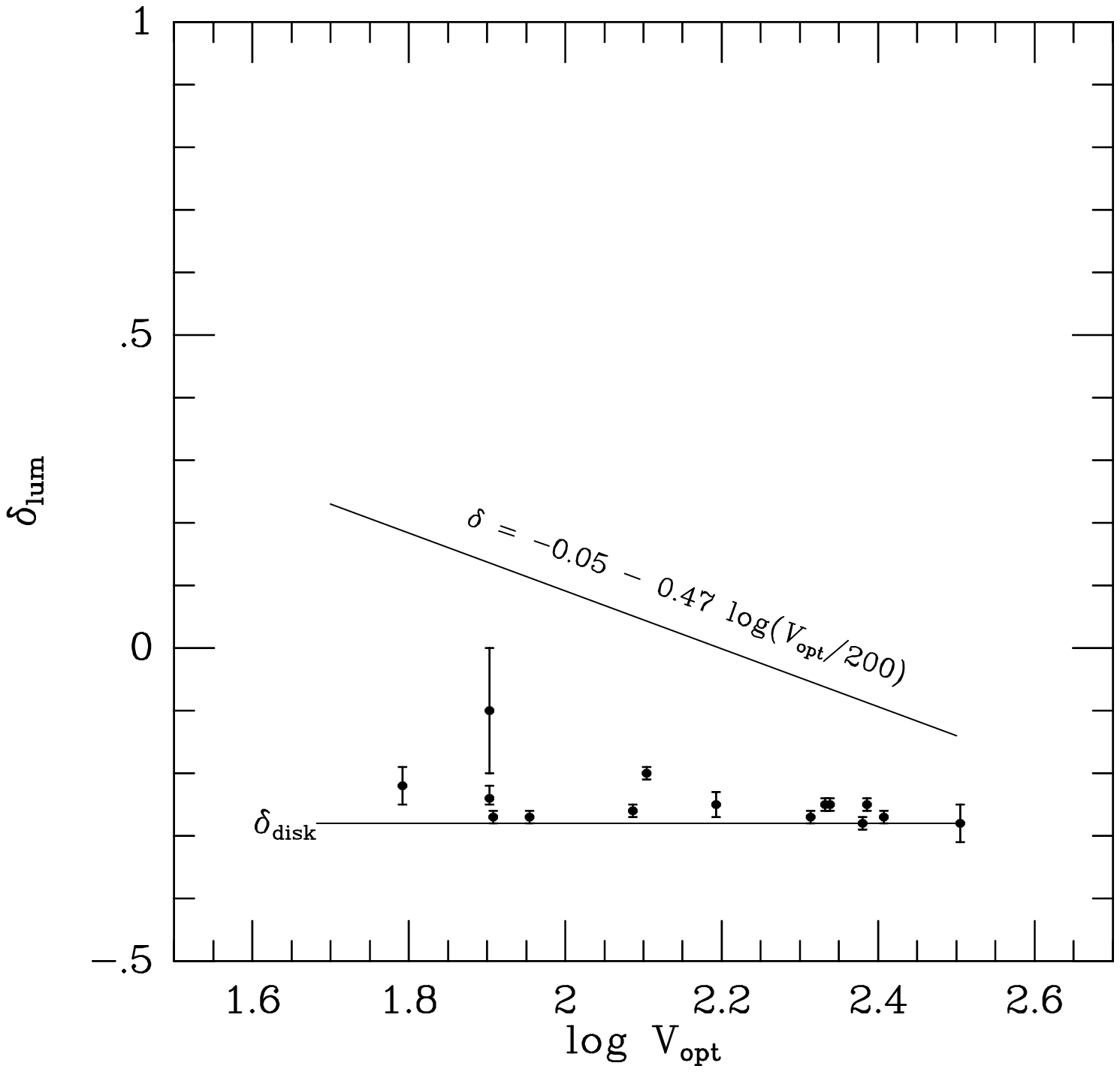,height=8cm,width=8cm}
}}
\par
\caption{ 
$\nabla_{lum}$ and $\delta_{lum}$ as a function of velocity. The data are
taken from Begeman (1987), Broeils (1992a,b), Carignan \& Puche (1990a,b),
Carignan, sancisi \& van Albada (1988), C\^ot\'e, Carignan \& Sancisi (1991), 
Puche, Carignan \& Bosma (1990), Puche, Carignan \& van Gorkom (1991b), and 
Puche, Carignan \& Wainscoat (1991a). The slanting and the horizontal line 
represent, respectively, the locus populated by the observed RC slopes and the 
exponential disc case.
}
\label{Figure A1}
\end{figure}
We can estimate $\nabla_{lum}$ from published data taking into account that
$$
\nabla_{lum}~ =
~ \biggl( {V_d \over V_{lum}} \biggr)^2 \beta_d ~+ ~
  \biggl( {V_b \over V_{lum}} \biggr)^2 \beta_b ~+ ~
  \biggl( {V_g \over V_{lum}} \biggr)^2 \beta_g
\eqno(A7)
$$
(where the subscripts d,b,g denote the disc, bulge and gas components, 
respectively), and 
$$
V_{lum}^2~ =~V_d^2 + V_b^2 + V_g^2\,.
\eqno(A8)
$$ 
In Fig.13 we plot $\nabla_{lum}$ as a function of luminosity: we find 
$$
\nabla_{lum} ~= ~-0.24 \pm 0.03 ~\simeq~ \dvd\,,
\eqno(A9)
$$
with the residuals uncorrelated with luminosity (but weakly correlated with Hubble 
type). This result can be easily understood. In fact, {\it (i)} most of the light 
profiles are true exponentials for which such a result holds automatically, {\it 
(ii)} the discrepancies of the surface brightness profiles with respect to an 
exponential are independent of luminosity, {\it (iii)} neglecting the bulge 
affects the determination of $\nabla$ very little (see Appendix of Persic et al. 
1993), and {\it (iv)} the gas content is relatively small, and its effect on 
$\nabla_{lum}$ is opposite to those considered in points {\it (ii)} and {\it 
(iii)}. 

The external gradient $\delta_{lum}$ can be estimated taking into account that 
$$
\delta_{lum} ~\simeq ~-0.26\,\beta_d ~ -~ 0.29\, \beta_b ~+ ~\beta_g
\delta_g\,. 
\eqno(A10)
$$

In Fig.13 we also plot $\delta_{lum}$ as a function of luminosity: we find that
there is very little dependence of this quantity on luminosity, so that we can
assume: 
$$
\delta_{lum}~ =~-0.25 \pm 0.04\,.
\eqno(A11)
$$

Since $\nabla$ and $\delta$ depend on luminosity while $\nabla_{lum}$ and 
$\delta_{lum}$ do not, then at least two of the halo quantities $1-\beta$, 
$\nabla_h$, $\delta_h$ will also depend on luminosity; on the other hand, a 
luminosity dependence of the latter quantities will generally be propagated 
into the RC profiles. 
\bigskip


\centerline{\bf APPENDIX B}

In this Appendix we show:

{\it a)} the rotation curves of Sample B co-added in velocity bins (Fig.14);

\begin{figure}

\par
\centerline{\psfig{figure=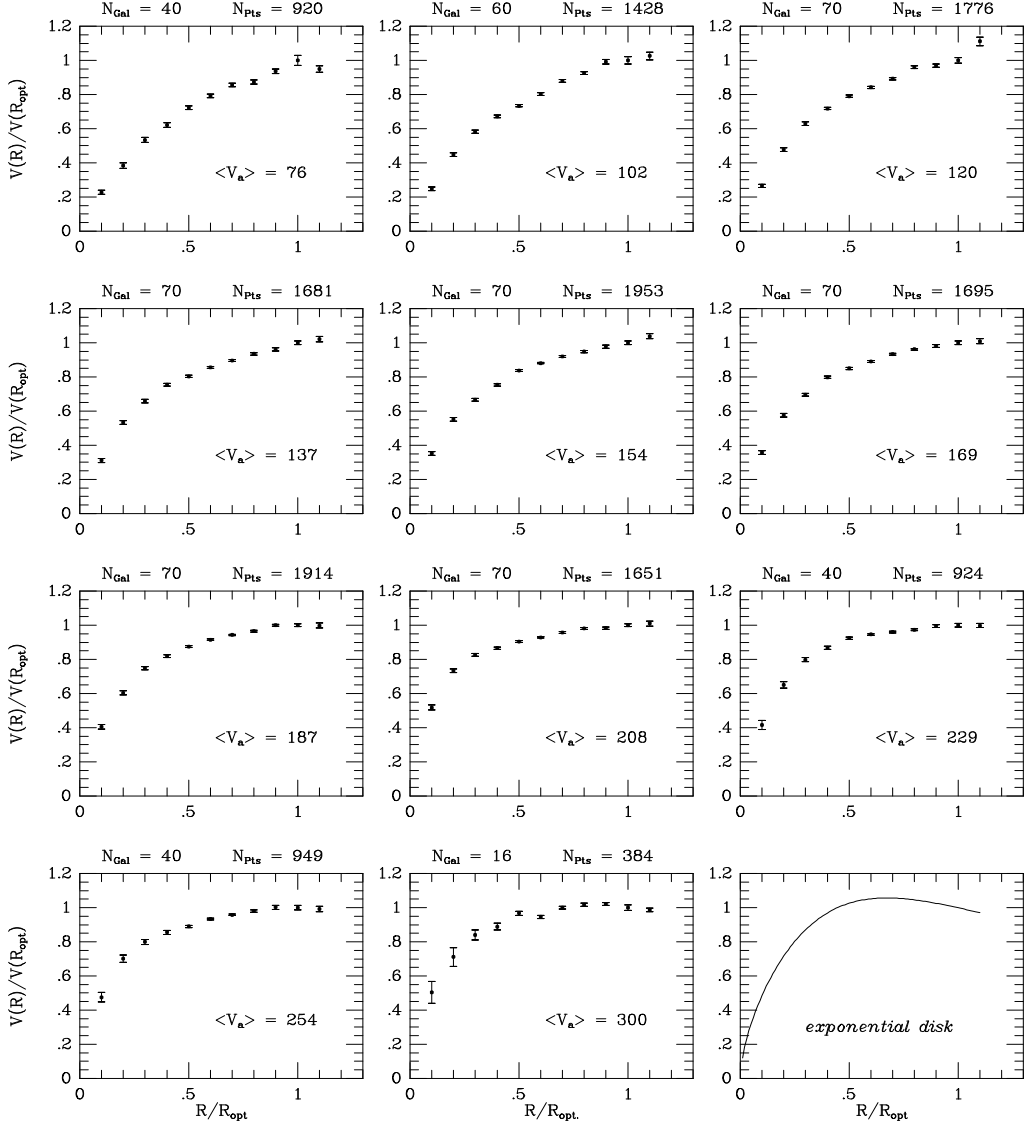,height=15cm,width=15cm}}
\par

\caption{ 
Synthetic rotation curves for Sample B arranged by velocity amplitude. 
Galactocentric radii are normalized to $R_{opt}$, the radius encompassing
$83\%$ of the total $I$ light. The last panel shows the rotation curve
predicted for a pure self-gravitating exponential thin disc.
}

\label{Figure B1}
\end {figure} 
{\it b)} the rotation curves of Sample B suitably co-added in order to probe
the kinematics of the outermost regions (Fig.15).
\bigskip
\begin{figure}
\par
\centerline{\psfig{figure=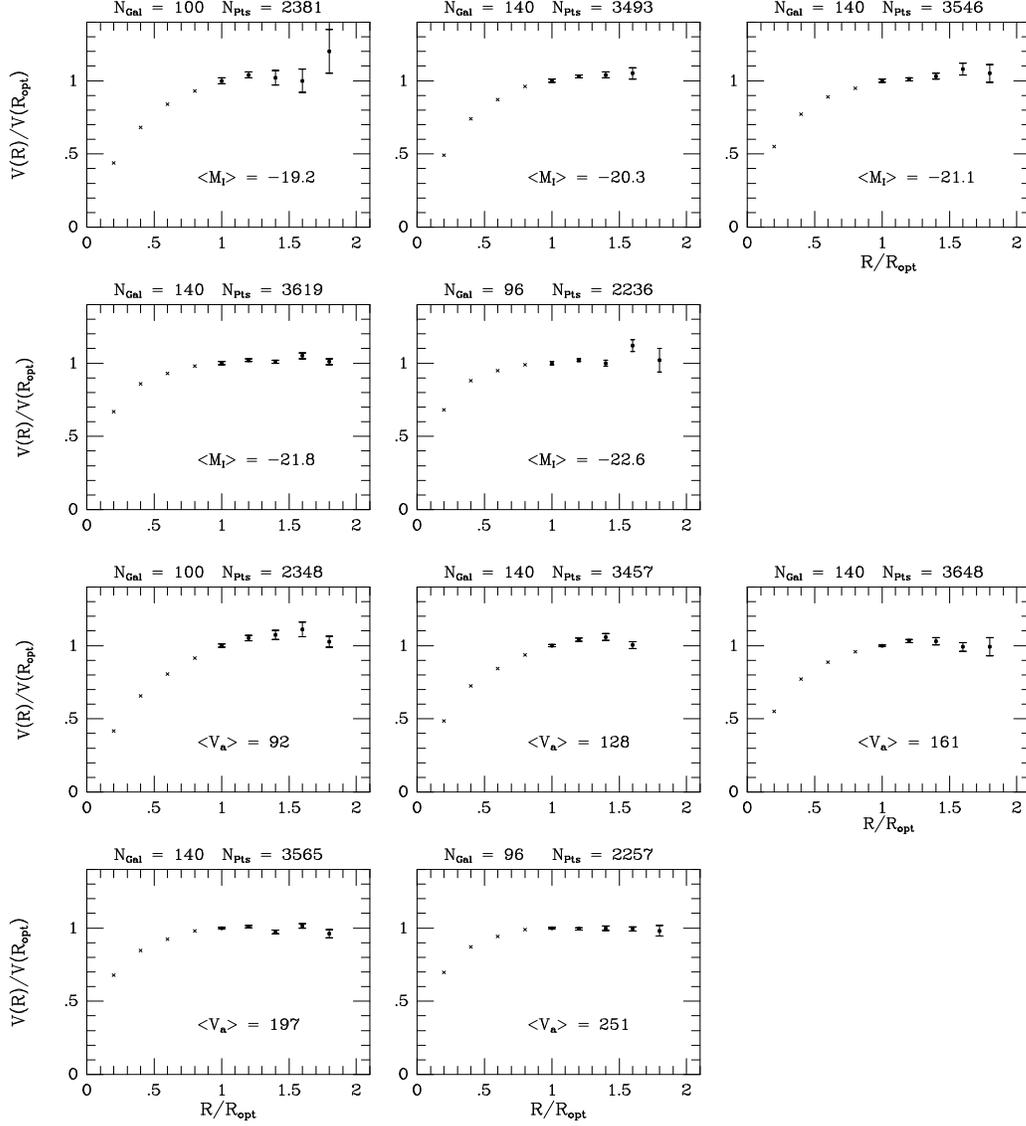,height=15cm,width=15cm}}
\par
\caption{
Luminosity sequence (upper five panels) and velocity-amplitude sequence 
(lower five panels) of the synthetic outer curves of Sample B (filled 
circles; the smaller symbols represent the inner curves, plotted with 
finer binning in Figs.1 and B1). The smaller number of luminosity/velocity 
bins is necessary to compensate for the sparseness of the outer data.
}
\label{Figure B2}
\end{figure}  

\begin{figure}
\par
\centerline{\psfig{figure=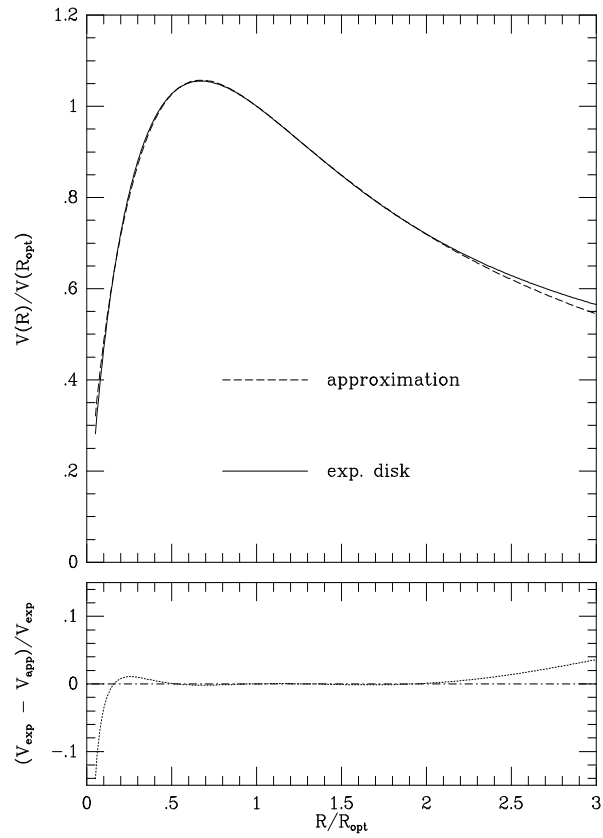,height=12cm,width=12cm}}
\par
\caption{ 
The approximated $V_d$ plotted alongside the exact function (upper panel)
The fractional difference is also shown (lower panel). 
}
\label{Figure C1}
\end{figure}
\centerline{\bf APPENDIX C}
\bigskip

The self-gravity of an infinitely thin disc with surface mass density distribution 
$I=I_0e^{R/R_D}$ yields an equilibrium  circular velocity given by:
$$
V^2_d(y)~ =~ {GM_{LM}\over 2\,R_D} ~ y^2~ (I_0K_0-I_1K_1) \,,
\eqno(C1) 
$$
where $y=R/R_D$ and $I_n$ and $K_n$ are the modified Bessel functions computed
at $y/2$. For mathematical simplicity, in the range $0.04 \leq x \equiv R/R_{opt} 
\leq 2$ we approximate the r.h.s of eq.(C1) with the much simpler function 
$$
V^2_d(R) ~= ~ 1.1 \, {GM_{LM} \over R_{opt}} ~ 
	{1.97~x^{1.22}\over ({x^2+0.78^2})^{1.43}}\,.
\eqno(C2)
$$ 
Notice in Fig.16 that our approximation agrees with the exact function to 
0.4\% over the quoted radial range. For $R>2R_{opt}$ the Keplerian regime is
practically attained, and hence: 
$$
V_d(R)~ =~V_d({2R_{opt}})~ \biggl( {R \over 2R_{opt}} \biggr)^{-1/2}\,.
\eqno(C3)
$$
\bigskip


\centerline{\bf APPENDIX D}

In this appendix we present the relevant quantities of Sample A (Table 1) along 
with the surface-photometry and rotational-velocity references (Table 2). 

\newpage

\par
\bigskip
\centerline{\hbox{
\psfig{figure=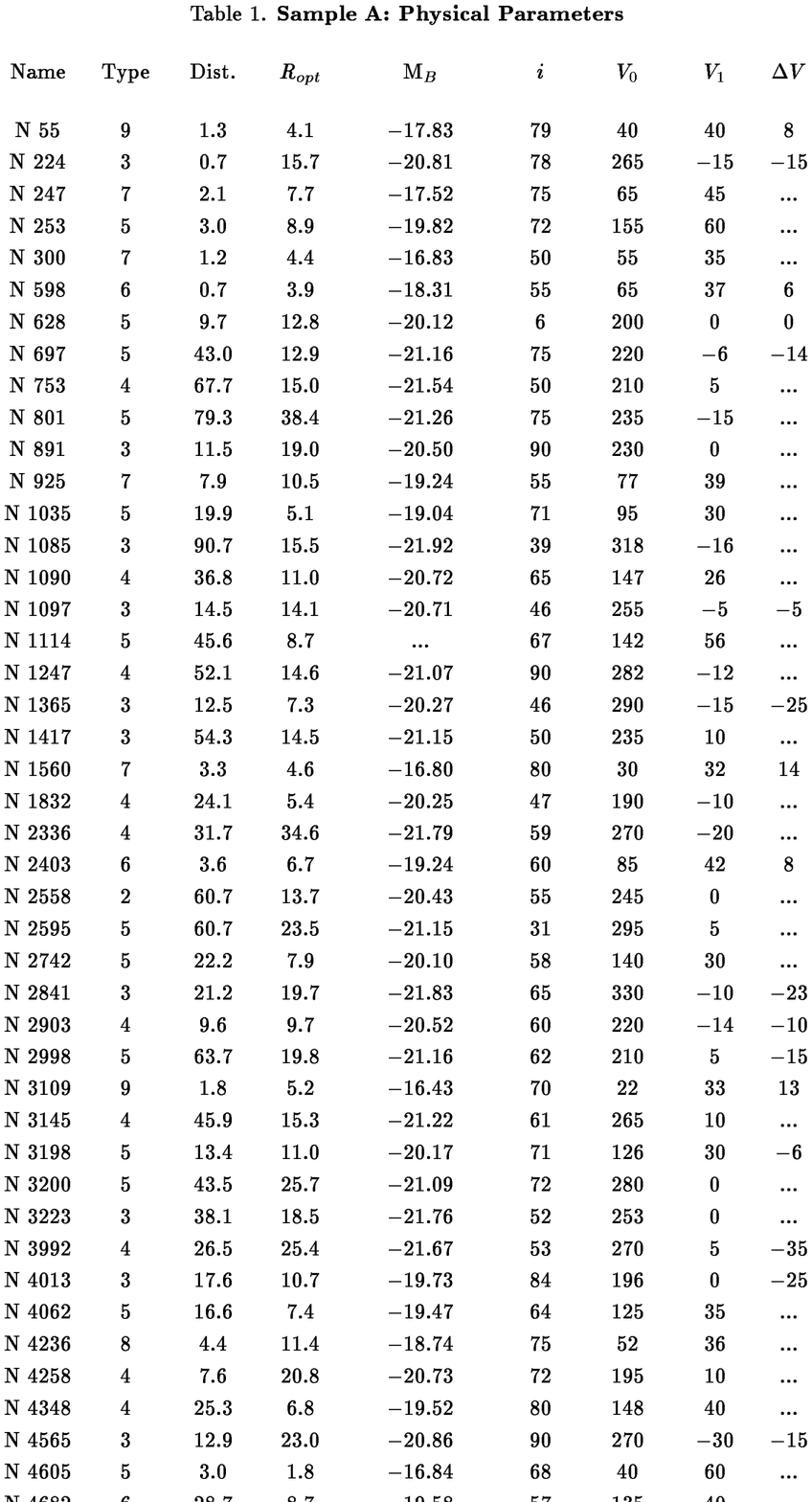,height=11cm,width=8cm}
\psfig{figure=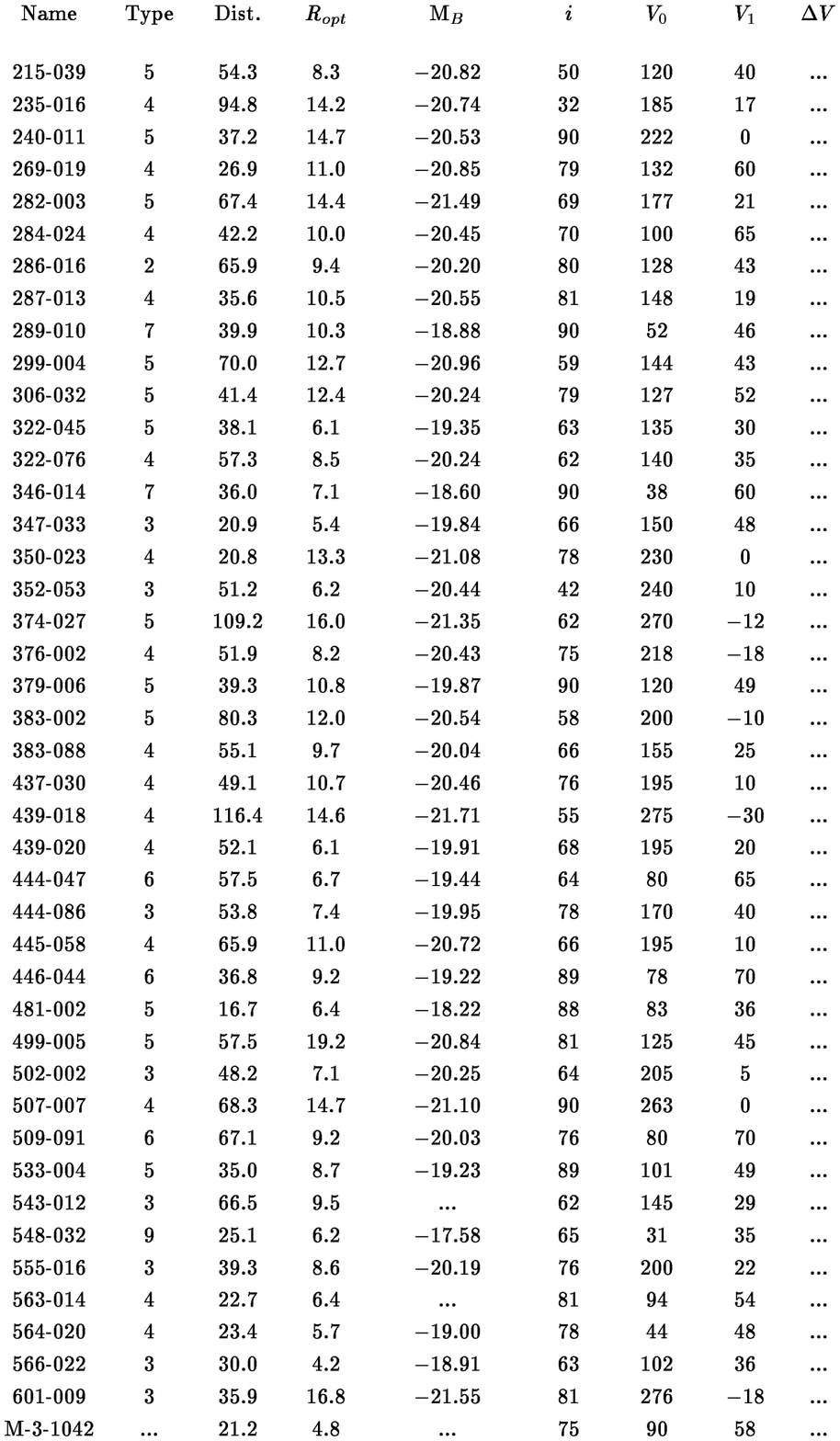,height=11cm,width=8cm}
}}

\medskip
\par
\centerline{\hbox{
\psfig{figure=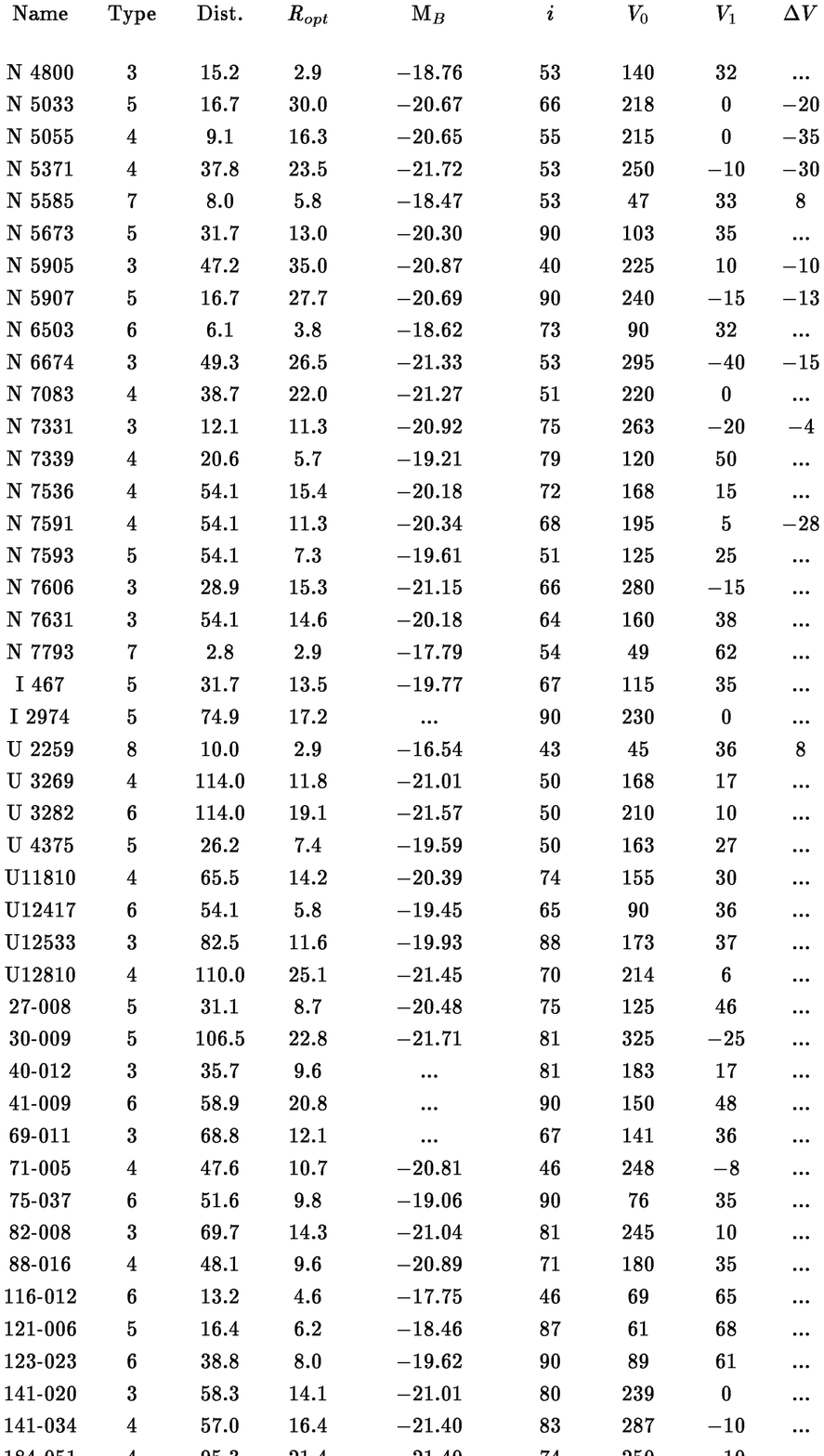,height=11cm,width=8cm}
\psfig{figure=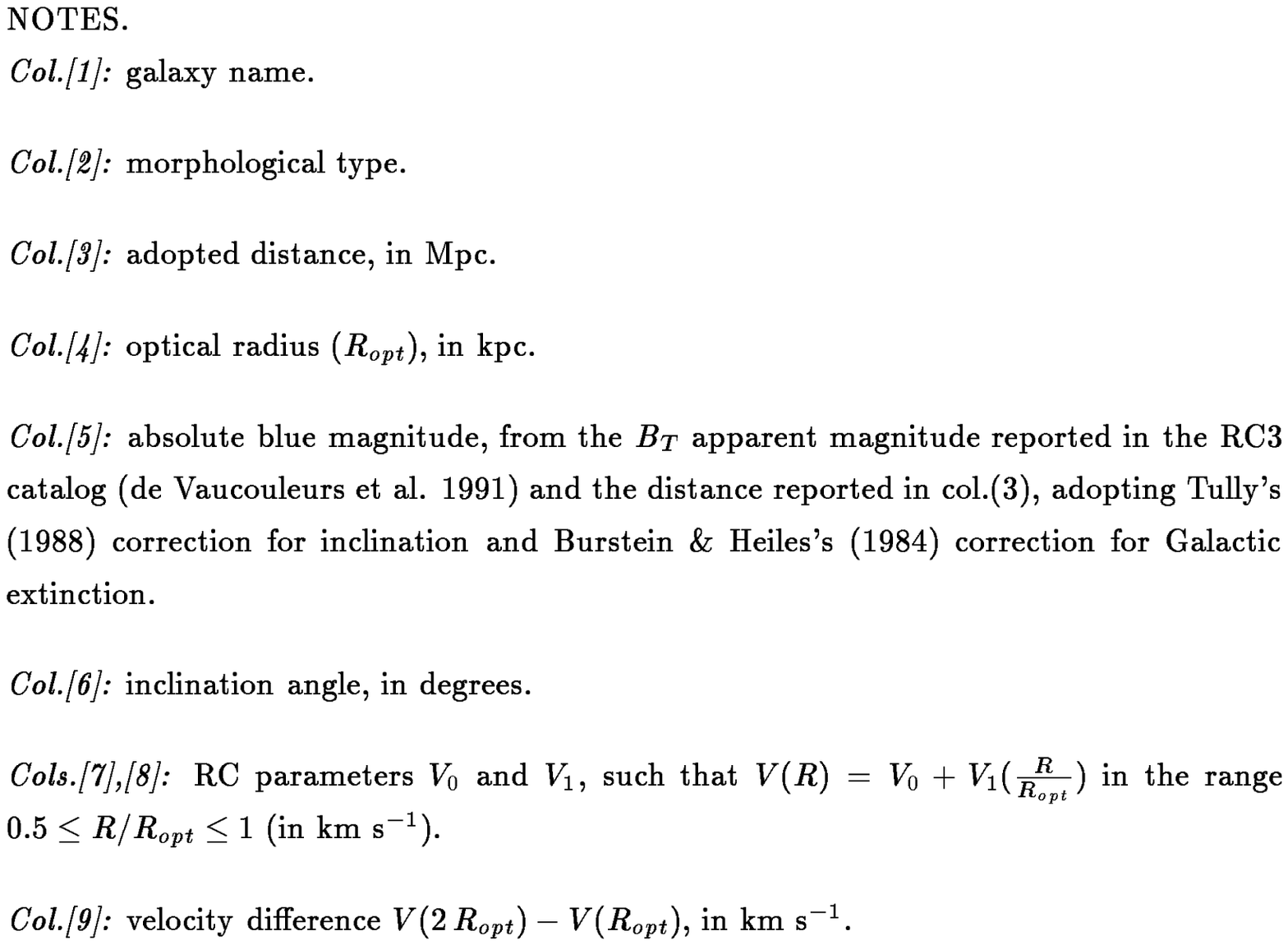,height= 3cm,width=8cm}
}}

\newpage
\par
\par
\centerline{\psfig{figure=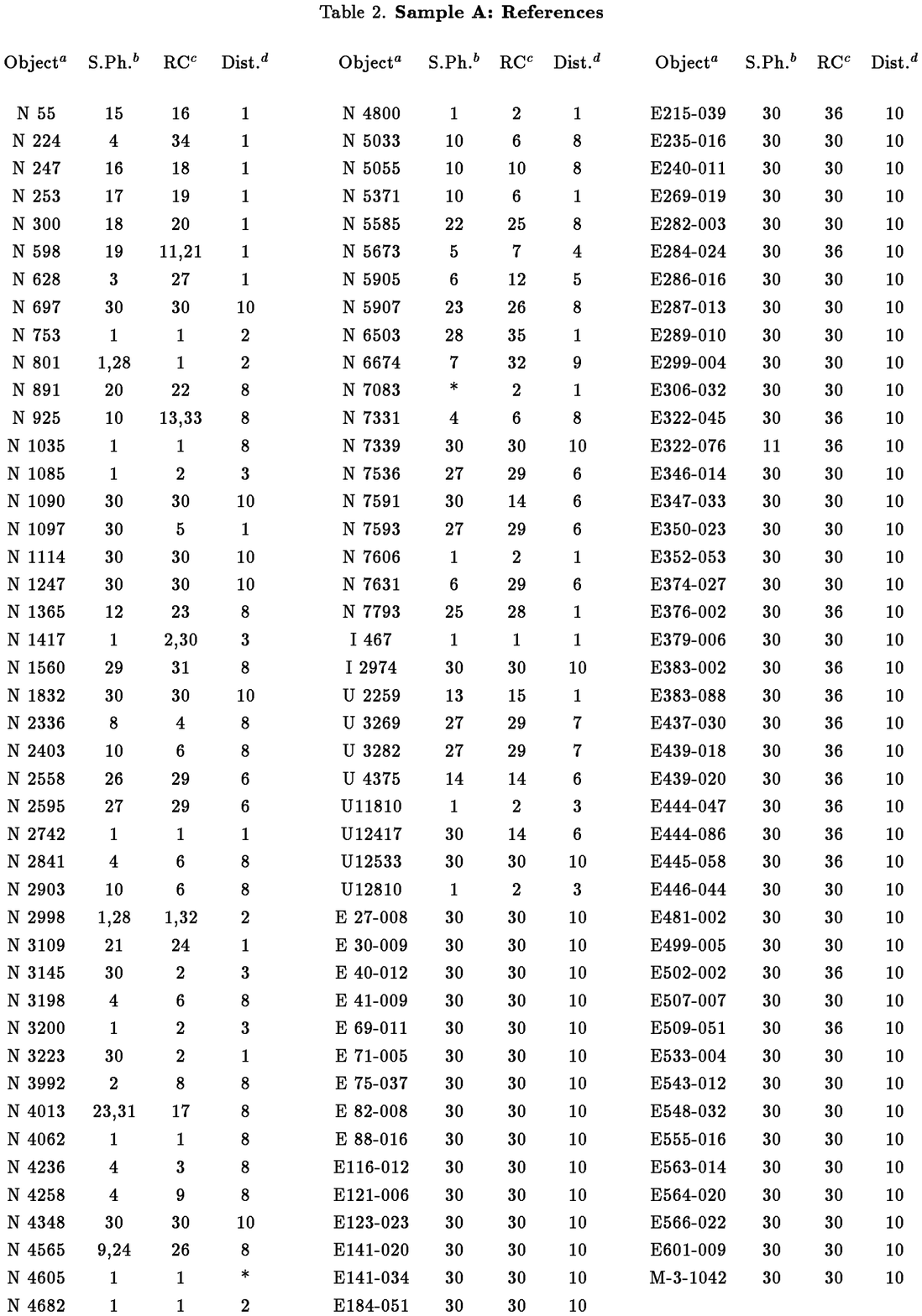,height=13cm,width=10cm}}
\centerline{\psfig{figure=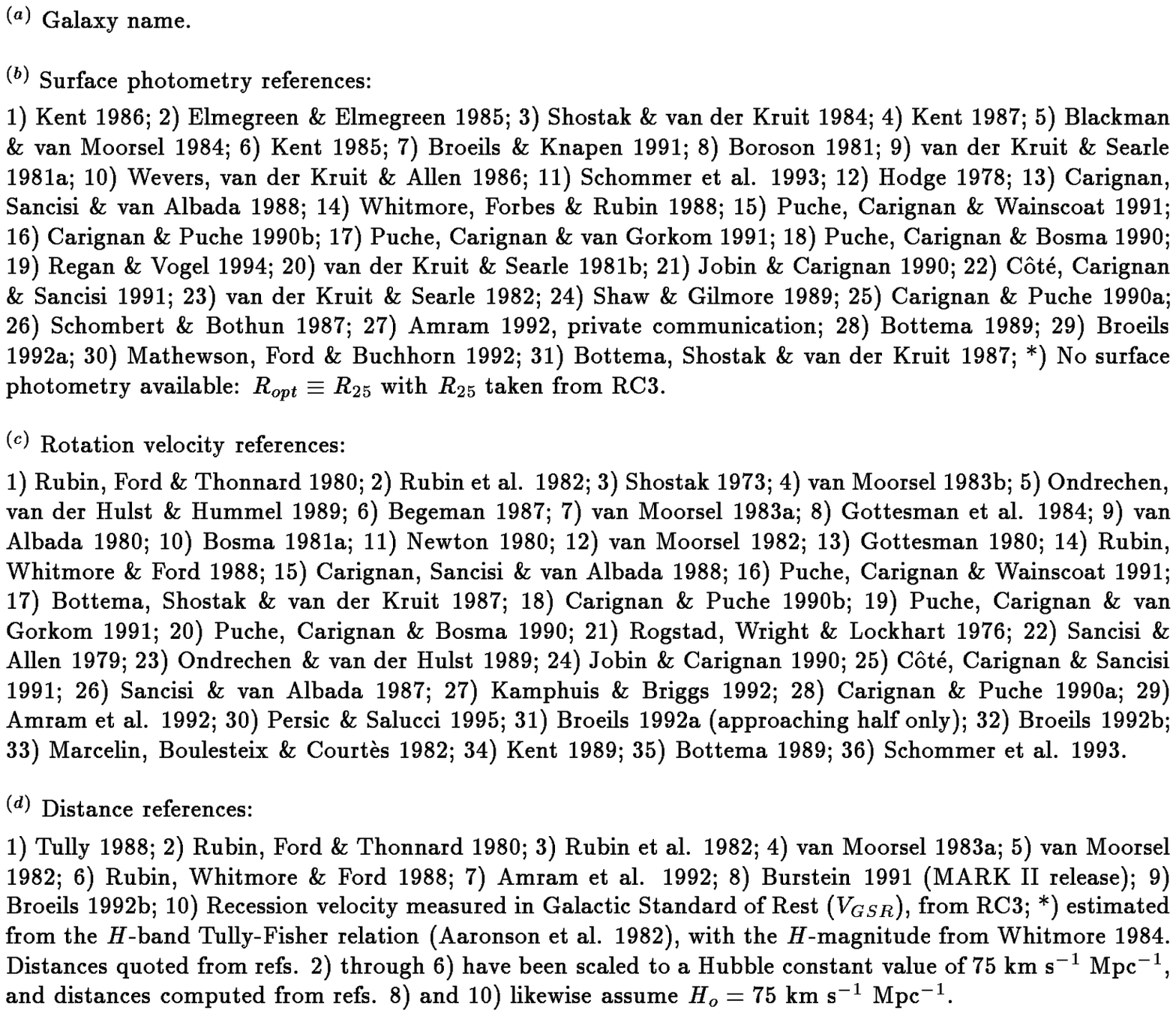,height= 8cm,width=10cm}}
\par 
\end{document}